\documentclass[%
reprint,
 aps,
prd,
floatfix,
superscriptaddress
]{revtex4-1}

\usepackage{graphicx}% Include figure files
\usepackage{dcolumn}% Align table columns on decimal point
\usepackage{bm}% bold math
\usepackage[mathlines]{lineno}% Enable numbering of text and display math
\usepackage{hyperref}
\newcommand{\commoncaption}{Top left: the outgoing muon momentum ($p_{\mu}$); Top right: the outgoing muon angle with respect to the initial neutrino direction ($\theta_{\mu}$); Bottom left: the outgoing pion angle with respect to the initial neutrino direction ($\theta_{\pi}$); Bottom right: the angle between outgoing muon and pion ($\theta_{\mu\pi}$).}

\begin{document}
\null\hfill\begin{tabular}[t]{l@{}}
%\small{Projects-doc-4628-v5}
\small{FERMILAB-PUB-18-142-ND}
\end{tabular}

\author{R.~Acciarri}
\affiliation{Fermi National Accelerator Lab, Batavia, Illinois 60510, USA}

\author{C.~Adams}
\affiliation{Yale University, New Haven, Connecticut 06520, USA}

\author{J.~Asaadi}
\affiliation{University of Texas at Arlington, Arlington, Texas 76019, USA}

\author{B.~Baller}
\affiliation{Fermi National Accelerator Lab, Batavia, Illinois 60510, USA}

\author{T.~Bolton}
\affiliation{Kansas State University, Manhattan, Kansas 66506, USA}

\author{C.~Bromberg}
\affiliation{Michigan State University, East Lansing, Michigan 48824, USA}

\author{F.~Cavanna}
\affiliation{Fermi National Accelerator Lab, Batavia, Illinois 60510, USA}
%\affiliation{Yale University, New Haven, Connecticut 06520, USA}

\author{E.~Church}
\affiliation{Pacific Northwest National Lab, Richland, Washington 99354, USA}

\author{D.~Edmunds}
\affiliation{Michigan State University, East Lansing, Michigan 48824, USA}

\author{A.~Ereditato}
\affiliation{University of Bern, 3012 Bern, Switzerland}

\author{S.~Farooq}
\affiliation{Kansas State University, Manhattan, Kansas 66506, USA}

\author{R.S.~Fitzpatrick}
\affiliation{University of Michigan, Ann Arbor, Michigan 48109, USA}

\author{B.~Fleming}
\affiliation{Yale University, New Haven, Connecticut 06520, USA}

% \author{H. Greenlee}
% \affiliation{Fermi National Accelerator Lab, Batavia, Illinois 60510, USA}

\author{A.~Hackenburg}
\affiliation{Yale University, New Haven, Connecticut 06520, USA}

% \author{R. Hatcher}
% \affiliation{Fermi National Accelerator Lab, Batavia, Illinois 60510, USA}

\author{G.~Horton-Smith}
\affiliation{Kansas State University, Manhattan, Kansas 66506, USA}

\author{C.~James}
\affiliation{Fermi National Accelerator Lab, Batavia, Illinois 60510, USA}

\author{K.~Lang}
\affiliation{University of Texas at Austin, Austin, Texas 78712, USA}

\author{I.~Lepetic}
\affiliation{Illinois Institute of Technology, Chicago, Illinois 60616, USA}

\author{B.R.~Littlejohn}
\affiliation{Illinois Institute of Technology, Chicago, Illinois 60616, USA}

\author{X.~Luo}
\affiliation{Yale University, New Haven, Connecticut 06520, USA}

\author{R.~Mehdiyev}
\affiliation{University of Texas at Austin, Austin, Texas 78712, USA}

\author{B.~Page}
\affiliation{Michigan State University, East Lansing, Michigan 48824, USA}

\author{O.~Palamara}
\affiliation{Fermi National Accelerator Lab, Batavia, Illinois 60510, USA}
%\affiliation{Yale University, New Haven, Connecticut 06520, USA}

% \author{G. Rameika}
% \affiliation{Fermi National Accelerator Lab, Batavia, Illinois 60510, USA}

\author{B.~Rebel}
\affiliation{Fermi National Accelerator Lab, Batavia, Illinois 60510, USA}

\author{A.~Schukraft}
\affiliation{Fermi National Accelerator Lab, Batavia, Illinois 60510, USA}

\author{G.~Scanavini}
\affiliation{Yale University, New Haven, Connecticut 06520, USA}

\author{M.~Soderberg}
\affiliation{Syracuse University, Syracuse, New York 13244, USA}

\author{J.~Spitz}
\affiliation{University of Michigan, Ann Arbor, Michigan 48109, USA}

\author{A.M.~Szelc}
\affiliation{University of Manchester, Manchester M13 9PL, United Kingdom}

\author{M.~Weber}
\affiliation{University of Bern, 3012 Bern, Switzerland}

\author{W.~Wu}
\affiliation{Fermi National Accelerator Lab, Batavia, Illinois 60510, USA}

\author{T.~Yang}
\email{tjyang@fnal.gov}
\affiliation{Fermi National Accelerator Lab, Batavia, Illinois 60510, USA}

\author{G.P.~Zeller}
\affiliation{Fermi National Accelerator Lab, Batavia, Illinois 60510, USA}

\collaboration{The ArgoNeuT Collaboration}
\noaffiliation

\begin{abstract}
We report on the first cross section measurement of charged-current single charged pion production by neutrinos and antineutrinos on argon. This analysis was performed using the ArgoNeuT detector exposed to the NuMI beam at Fermilab. The measurements are presented as functions of muon momentum, muon angle, pion
angle, and angle between muon and pion. The flux-averaged cross sections are measured to be
$2.7\pm0.5(stat)\pm0.5(syst) \times 10^{-37} \textrm{cm}^{2}/\textrm{Ar}$ for neutrinos at a mean energy of 9.6 GeV and $8.4\pm0.9(stat)^{+1.0}_{-0.8}(syst) \times 10^{-38} \textrm{cm}^{2}/\textrm{Ar}$ for antineutrinos at a mean energy of 3.6 GeV with the charged pion momentum above 100 MeV/$c$. The results are compared with several model predictions.
%The integrated cross sections are measured to be $xx\pm yy\textrm{(stat)}^{z1}_{z2}\textrm{(syst)}\times 10^{-38}\textrm{cm}^2/\textrm{nucleon}$ for neutrinos at a mean energy of 9.6 GeV and $xx\pm yy\textrm{(stat)}^{z1}_{z2}\textrm{(syst)}\times 10^{-38}\textrm{cm}^2/\textrm{nucleon}$ for antineutrinos at a mean energy of 3.6 GeV. The differential cross sections are compared with 
%\begin{description}
%You may use the \texttt{description} environment to structure %your abstract;
%use the optional argument of the \verb+\item+ command to give %the category of each item. 
%\end{description}
\end{abstract}

\title{First measurement of the cross section for $\nu_\mu$ and $\bar{\nu}_\mu$ induced single charged pion production on argon using ArgoNeuT}% Force line breaks with \\

% \altaffiliation[Also at ]{Physics Department, XYZ %University.}%Lines break automatically or can be forced with \\
%\affiliation{%

%\collaboration{ArgoNeuT Collaboration}%\noaffiliation

\date{\today}% It is always \today, today,
\maketitle

\maketitle

%\tableofcontents

\section{\label{sec:introduction}Introduction}

Precision neutrino cross section measurements are crucial in order to fully characterize the properties of the neutrino-nucleus interaction. In present and future neutrino oscillation experiments, such as T2K, NOvA, SBN, DUNE, and HyperKamiokande~\cite{Abe:2011ks,Ayres:2007tu,Antonello:2015lea,dune,Abe:2015zbg}, the neutrino-nucleus interaction must be well understood in order to reconstruct the incoming neutrino from properties of the final state.  

Charged-current single pion production (CC1$\pi^{\pm}$) is an important process in few-GeV neutrino-nucleus interactions~\cite{Formaggio:2013kya}. The process is mainly produced through nucleon resonances and deep inelastic scattering. The earliest neutrino CC1$\pi^{\pm}$ measurements used hydrogen or deuterium targets. There are some discrepancies among this early data, most notably, the ANL~\cite{Radecky:1981fn} and BNL~\cite{Kitagaki:1986ct} measurements differ by up to ∼40\% in normalization. A recent reanalysis of the two experiments prefers the ANL measurement~\cite{Wilkinson:2014yfa}. The nuclear medium plays an important role in the production and propagation of hadrons produced in neutrino-nucleus interactions. Hadrons produced in neutrino-nucleus interactions may re-scatter while propagating through the nuclear medium, referred to as final-state interactions (FSI), and can change the charge and multiplicity of the outgoing hadrons, as well as altering their kinematics~\cite{Dytman:2009zza, Golan:2012wx}. There are more modern measurements of $\nu_{\mu}$ CC1$\pi^{+}$ on various nuclear targets which show varying degrees of agreements with different model predictions. The MiniBooNE~\cite{AguilarArevalo:2010bm}, MINER$\nu$A~\cite{Eberly:2014mra} and T2K~\cite{Abe:2016aoo} Collaborations provided measurements of this cross section in mineral oil, plastic scintillator and water, respectively.

We present the first CC1$\pi^{\pm}$ differential cross section measurement on argon using the ArgoNeuT (Argon Neutrino Test) detector~\cite{Anderson:2012vc}. The signal is defined to be a charged-current $\nu_{\mu}$ or $\bar{\nu}_{\mu}$ interaction in the detector, with one charged pion above 100 MeV/$c$ momentum exiting the target nucleus. Events with neutral or charged kaons or neutral pions or more than one charged pions above 100 MeV/$c$ momentum in the final state are excluded. The strategy adopted in this analysis relies on classifying events, after a chain of topology cuts, with the help of the ROOT Toolkit for Multivariate Analysis~\cite{tmva}, with which a boosted decision tree (BDT) is created using reconstructed quantities in the TPC fiducial volume. 

The results presented here will serve to help better constrain the modeling of neutrino-nucleus interactions. In practice, the ArgoNeuT CC1$\pi^{\pm}$ cross section measurement will provide useful information on the single pion production to the Monte Carlo (MC) generators, improving the constraints on both resonant pion production and FSI models.  These data will be of particular benefit to the planned DUNE experiment~\cite{Acciarri:2016crz}, which will use argon as the target in its far detector.

\section{\label{sec:detector}ArgoNeuT Experiment}
ArgoNeuT is a 170 L liquid argon time projection chamber (LArTPC) with dimensions $47\times40\times90$ cm$^3$ (horizontal drift dimension $\times$ height $\times$ length). The electric field inside the TPC is 481 V/cm, and the drifted charge from particle interactions is read out in two planes of 240 wires each (the induction and collection planes) with a plane spacing and a wire pitch of 4 mm. The angle between the induction and collection plane wires is 60 degrees. 

ArgoNeuT collected neutrino and antineutrino events in Fermilab's NuMI beam line~\cite{Adamson:2015dkw} at the MINOS near detector hall from September, 2009 through February, 2010. In combination with ArgoNeuT, the downstream MINOS near detector is used to fully reconstruct neutrino-induced events. The measurements reported in this paper are based on data taken with the NuMI beam line operating in the reverse horn current (antineutrino enhanced) mode, corresponding to $1.25\times10^{20}$ protons on target (POT) during which both ArgoNeuT and MINOS near detector were operational. 60\% of all interactions in ArgoNeuT derive from neutrinos, while the remaining 40\% are produced by antineutrinos~\cite{Adamson:2013ue}. The flux-averaged neutrino energy is 9.6 GeV for neutrinos and 3.6 GeV for antineutrinos.

The MINOS near detector is a 1 kton magnetized steel/scintillator tracking/sampling calorimeter~\cite{Michael:2008bc} and it operated approximately 1.5 m downstream of ArgoNeuT. The muons that exit ArgoNeuT's TPC volume are matched to MINOS, where the momentum and charge are reconstructed. ArgoNeuT is not magnetized so there is no possibility to separate pions according to their charge.

\begin{figure}[!ht]
\centering
\includegraphics[width=0.45\textwidth]{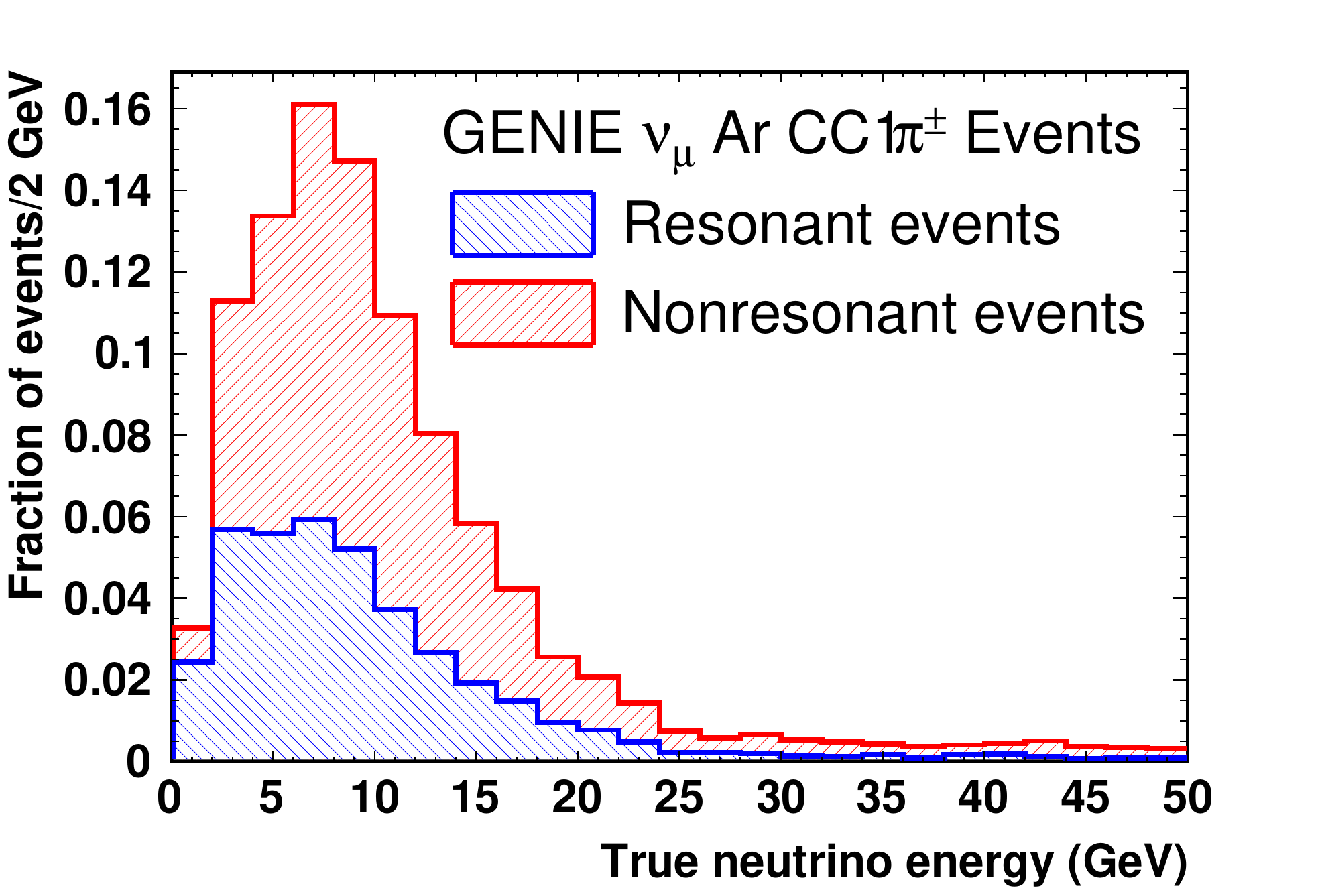}
\includegraphics[width=0.45\textwidth]{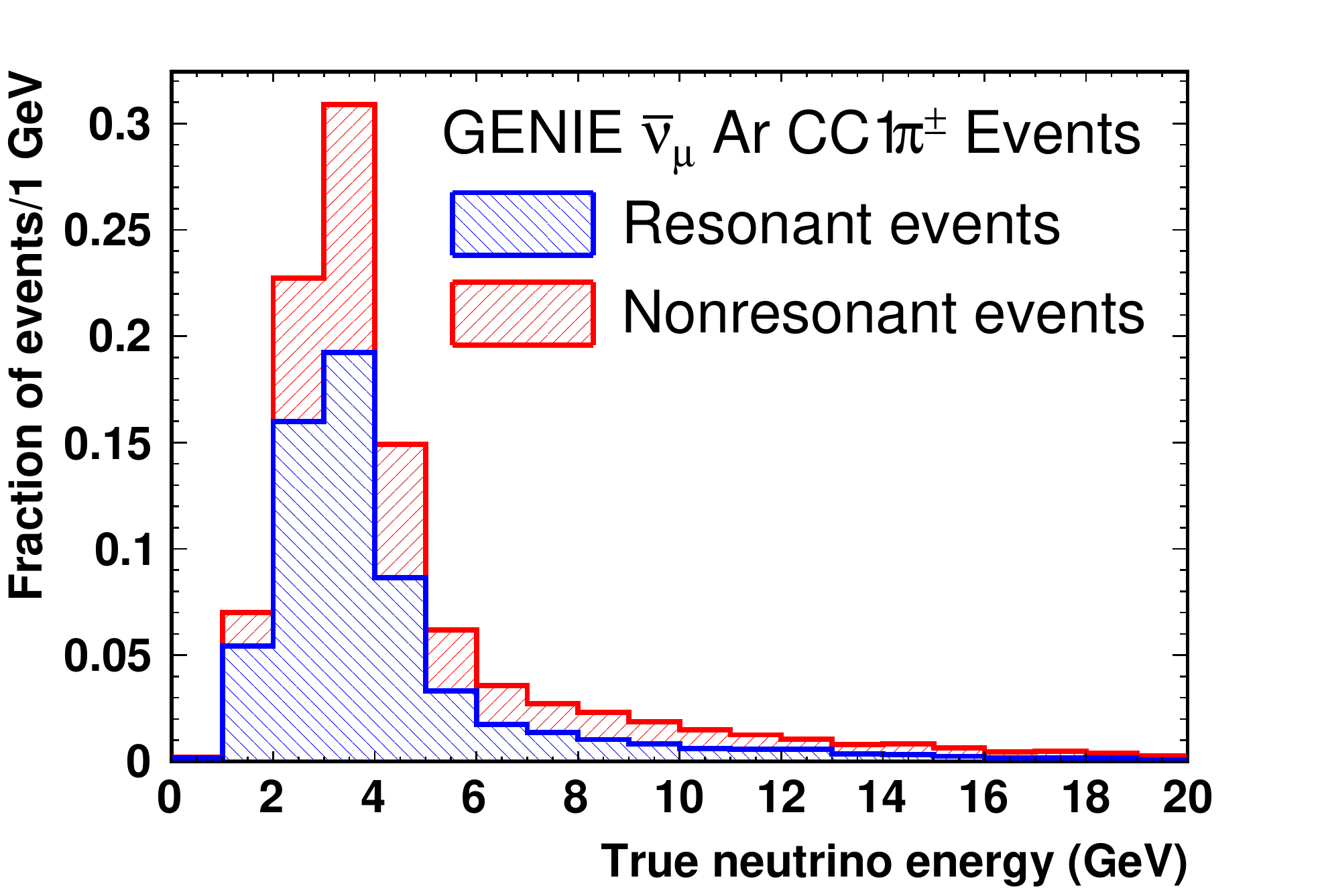}
\caption{Stacked compositions of resonance events and non-resonance events in simulated ArgoNeuT CC1$\pi^{\pm}$ events. The top figure is for the neutrino-induced events while the bottom figure is for the antineutrino-induced events.}\label{fig:res}
\end{figure}

Figure~\ref{fig:res} shows the resonant pion production and non-resonant pion production components in simulated ArgoNeuT CC1$\pi^{\pm}$ events using the GENIE~\cite{Andreopoulos:2009rq} neutrino event generator. Nonresonant pion production involves quasielastic scattering, deep inelastic scattering (nonresonant multi-hadron production), coherent pion production and meson exchange current events. According to the GENIE generator, resonant pion production contributes 39\% and 61\% to the $\nu_{\mu}$ and $\bar{\nu}_{\mu}$ CC1$\pi^{\pm}$ cross sections, respectively. The remaining CC1$\pi^{\pm}$ events are produced predominantly through deep inelastic scattering. 89\% of the $\nu_{\mu}$ CC1$\pi^{\pm}$ events contain a $\pi^{+}$ while 97\% of the $\bar{\nu}_{\mu}$ CC1$\pi^{\pm}$ events contain a $\pi^{-}$. Coherent pion production contributes 3\% and 5\% to the $\nu_{\mu}$ and $\bar{\nu}_{\mu}$ CC1$\pi^{\pm}$ cross sections, respectively~\cite{Acciarri:2014eit}.

\section{\label{sec:ana}Event Reconstruction and Selection}

The drifting of ionization electrons in liquid argon and the signal digitization in the ArgoNeuT detector are simulated in LArSoft~\cite{larsoft}. The neutrino and antineutrino interactions inside ArgoNeuT are reconstructed using the LArSoft software package, rendering a full characterization of the charged particles emerging in the ArgoNeuT detector.  The reconstruction software provides the spatial information of each reconstructed track as well as the particle identification based on the calorimetric information. The details of the reconstruction chain used by this analysis are described in Appendix~\ref{sec:reco}. 

After tracking final state muons in ArgoNeuT, an attempt is made to match the 3D tracks that leave ArgoNeuT with muons that have been reconstructed in MINOS and have a hit within 20 cm of the upstream face of the MINOS detector. This matching criterion is based on the radial and angular differences between the projected-to-MINOS ArgoNeuT track and the candidate MINOS track, taking into account the expected muon multiple scattering. In the ArgoNeuT detector, the muon energy loss before reaching MINOS is calculated from the distance the muon traverses in liquid argon. A stand-alone version of MINOS simulation and reconstruction is used to characterize the tracks passing from ArgoNeuT into MINOS. If the muon is reconstructed in the MINOS near detector, the muon momentum is measured from its range or the track curvature, depending on whether the muon stops in MINOS or not. 

\begin{figure}[!ht]
\centering
\includegraphics[width=0.45\textwidth]{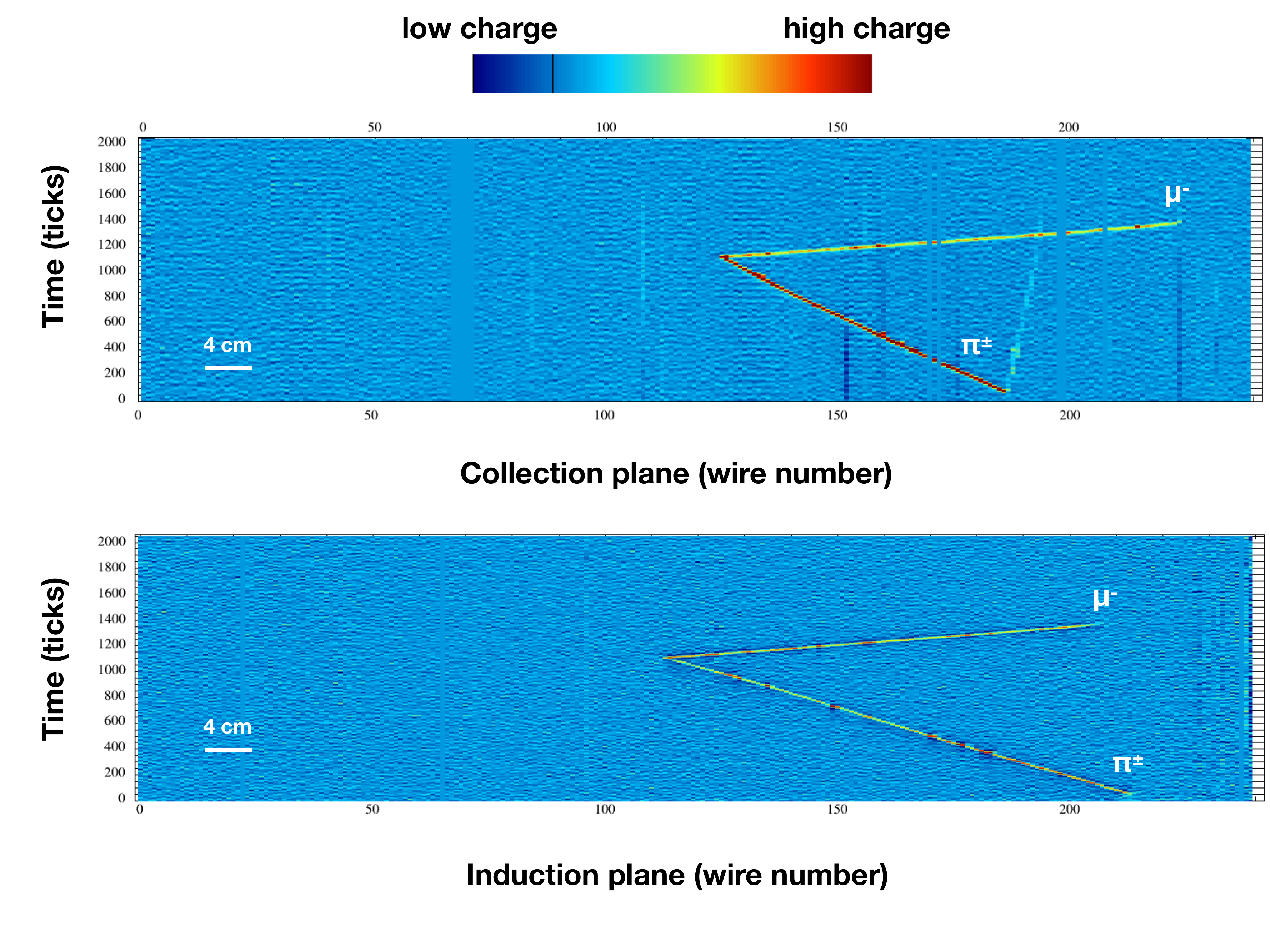}
\includegraphics[width=0.45\textwidth]{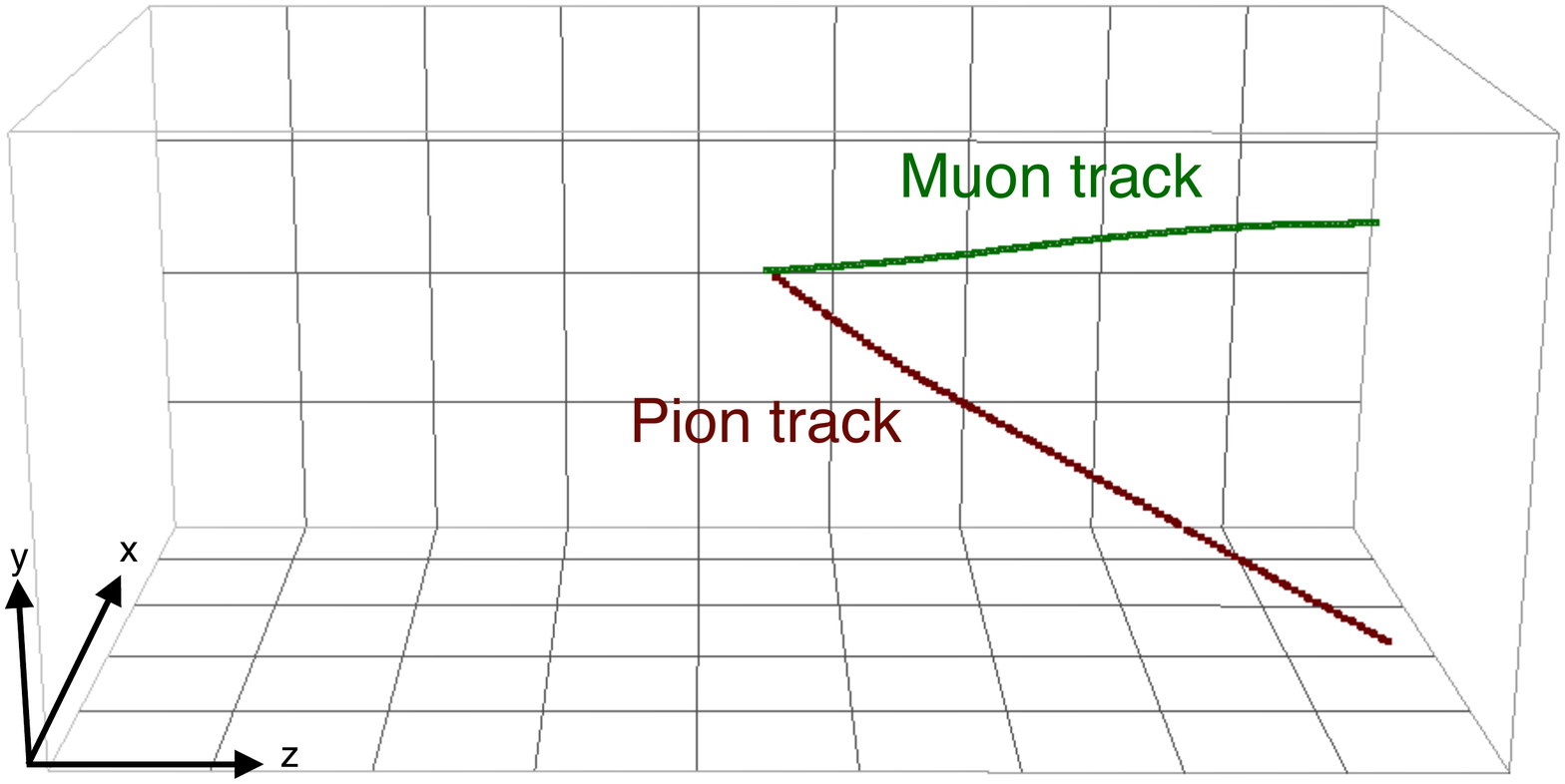}
\caption{A $\mu^{-}\pi^{\pm}$ candidate event in ArgoNeuT. The top figure shows the 2D projection in the two wire planes while the bottom figure shows the 3D reconstructed tracks. }\label{fig:evd}
\end{figure}

Fig.~\ref{fig:evd} shows a $\mu^{-}\pi^{\pm}$ candidate event in ArgoNeuT. The LArSoft reconstruction package successfully reconstructs both charged particles as 3D tracks, with both tracks exiting the LArTPC and the most forward going track matched to a track in the MINOS detector. 

In this analysis, we first apply the $\nu_{\mu}$/$\bar{\nu}_{\mu}$ CC event selection that is similar to what we developed in the previous ArgoNeuT CC-inclusive analyses~\cite{Anderson:2011ce, Acciarri:2014isz}. We require that there is at least one track in ArgoNeuT that is matched to a track in the MINOS detector. The matched track in ArgoNeuT is identified as the ``muon track''. All the reconstructed vertices within 4 cm of the start point of the muon track are considered and the closest vertex to the start point of the muon track is identified as the neutrino interaction vertex. The reconstructed neutrino interaction vertex is required to be inside of the ArgoNeuT fiducial volume, defined as a rectangular box shaped as follows: the boundary from the plane comprising the induction plane and that of the cathode plane is 3 cm, the boundary from the top and bottom of the TPC is 4 cm, and the distances from the upstream and downstream ends are 6 cm and 4 cm, respectively.

Neutrino-induced through-going muons that pass the fiducial volume requirement due to possible inefficiency in the vertex reconstruction are removed by two additional requirements. First, we reconstruct all the possible 3D points using all hit pairs on the two wire planes that have the same drift time. We reject events that have any 3D points within 4 cm to the LArTPC front surface. Secondly, we consider all the uncontained tracks that start within 4 cm of the reconstructed neutrino interaction vertex and calculate the angle between each track and the muon track. One major background arises from the presence of broken tracks for which, due to a reconstruction failure, a single through-going muon is reconstructed as two tracks going back-to-back. We reject events where the largest angle between the muon track and any other track is above 170 degrees, which indicates a through-going muon is broken into two tracks.

Our goal is to detect events with one muon track, one charged pion track, and any number of protons. After the removal of through-going muons, we investigate all tracks that are at least 4 cm long and start within 4 cm of the reconstructed neutrino interaction vertex, excluding the muon track matched to MINOS. The track length requirement imposes an approximate 100 MeV/$c$ threshold on the pion momentum. If only one track is identified as charged pion or muon (stopping charged pions and muons are almost indistinguishable in a LArTPC) by the calorimetry based particle identification, detailed in Appendix~\ref{sec:reco}, the event is selected.  Because of the small size of ArgoNeuT, many protons are not contained in the detector and they are often flagged as charged pion or muon using only the measured $dE/dx$ information inside the TPC. As a consequence, approximately 16\% of the true CC1$\pi^{\pm}$ signal events have two tracks that are identified as charged pion or muon in addition to the muon track. We relax the selection to also accept events with one muon track matched to MINOS and two tracks identified as charged pion or muon to improve the signal selection efficiency. We take track identified as charged pion or muon with the lower average $dE/dx$ (energy loss per unit length) as the charged pion candidate. %Based on the MC studies, ??\% of the selected charged pion track is produced by a real charged pion for the CC1$\pi^{\pm}$ signal events.

To further improve the signal to background ratio, we created boosted decision trees (BDT) using the ROOT Toolkit for Multivariate Analysis. The BDT is trained using simulated signal and background samples. The simulation of neutrino interactions in ArgoNeuT employs a {\sc geant}4-based~\cite{Agostinelli:2002hh} detector model and particle propagation software in combination with the GENIE neutrino event generator. The neutrino and antineutrino fluxes for the NuMI beam were taken from reference~\cite{Aliaga:2016oaz}, by the MINER$\nu$A Collaboration. To minimize the dependence on a particular neutrino generator model, we only use the calorimetry and event topology information when we build the BDT. The input variables are:
\begin{itemize}
\item Average $dE/dx$ calculated using the last 5 cm of the candidate pion track;
\item Number of tracks identified as charged pion or muon;
\item Number of reconstructed vertices;
\item Fraction of total measured charge that is associated with all reconstructed tracks;
\item Fraction of total measured charge that is associated with reconstructed tracks originating from the reconstructed neutrino interaction vertex. 
\end{itemize}

The last three variables are chosen to remove the deep inelastic scattering background where the complex topology tends to lead to many reconstructed vertices and tracks or failure in reconstructing some tracks. Two BDTs are trained for neutrino and antineutrino samples separately. Table~\ref{tab:numevt} shows the number of data events after each selection requirement. More detailed information can be found in \cite{Scanavini:2017trd}.

\begin{table}[!ht]
\caption{Number of data events after each selection requirement. Events with BDT score $>$ 0 are considered as signal candidates.}\label{tab:numevt}
\centering
\begin{tabular}{ c c c }
\hline
& $\nu_{\mu}$ & $\bar{\nu}_{\mu}$ \\
\hline
$\nu_{\mu}$/$\bar{\nu}_{\mu}$ CC selection & 1862 & 1756 \\
One or two MIP-like tracks & 907 & 624 \\
BDT score $>$ 0 & 337 & 285 \\
BDT score $<$ 0 & 570 & 339 \\
\hline
\end{tabular}
\end{table}

\section{\label{sec:eff}Background estimation, unfolding and cross section results}

\begin{figure*}[!ht]
\centering
\includegraphics[width=0.45\textwidth]{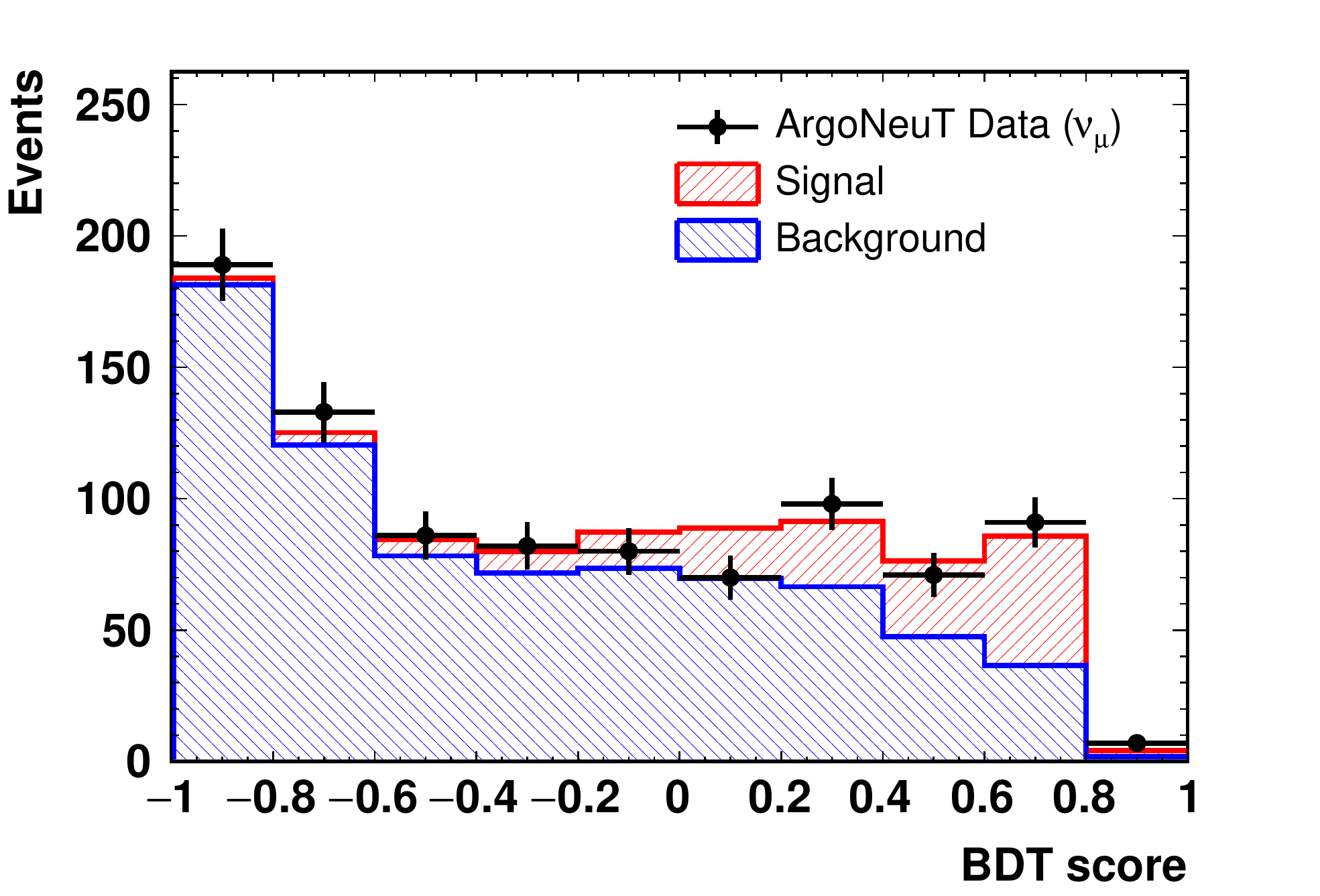}
\includegraphics[width=0.45\textwidth]{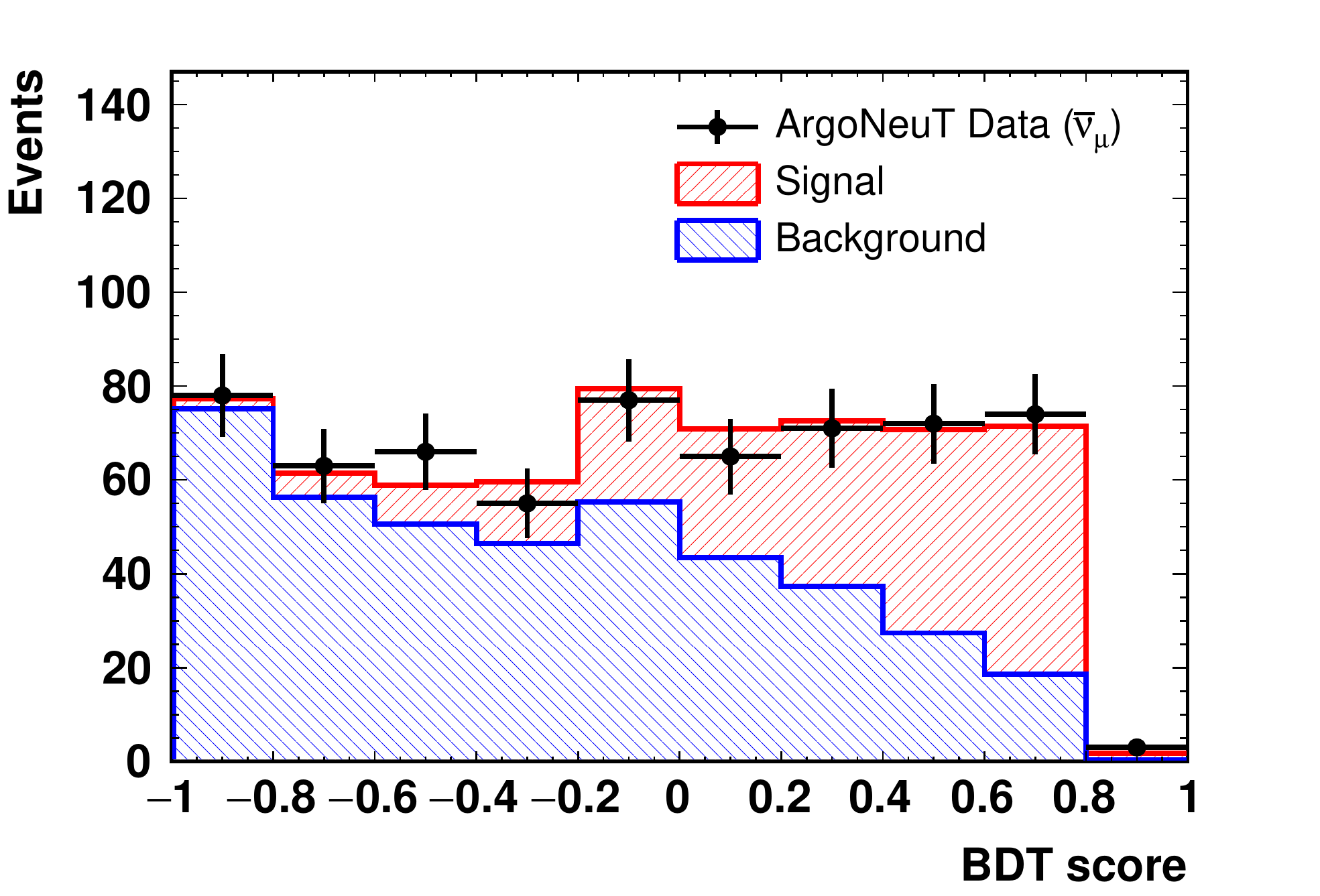}
\caption{The fits of the BDT distributions in data to simulated (GENIE) CC1$\pi^{\pm}$ signal and background. The left figure is for the neutrino events while the right figure is for the antineutrino events.}\label{fig:bdtfit}
\end{figure*}

To estimate the rate of background events in the signal region (BDT score $>$ 0), the BDT distributions in data for $\mu^{-}$ and $\mu^{+}$ events are fit to a linear combination of templates for CC1$\pi^{\pm}$ signal and background, obtained from a simulated neutrino and antineutrino sample using the GENIE neutrino event generator, which is the same method used in Ref.~\cite{Acciarri:2014eit}. Here only the shape information of the simulated BDT distributions is used, which helps to reduce the systematic uncertainties. Fig.~\ref{fig:bdtfit} shows the fits of the BDT distributions. Table~\ref{tab:bdtfit} summarizes the results of the fits. We consider data events with BDT $>$ 0 as CC1$\pi^{\pm}$ signal candidates and the fitted number of background events in the region BDT $>$ 0 is subtracted from data.
\begin{table*}[!ht]
\centering
\caption{Results of BDT fits}\label{tab:bdtfit}
\begin{tabular}{c c c c c|c c c c}
\hline
&\multicolumn{4}{c}{$\nu_\mu$} & \multicolumn{4}{c}{$\bar{\nu}_\mu$}\\
\hline
 & Data & Fitted MC & Fitted Signal & Fitted Background & Data & Fitted MC & Fitted Signal & Fitted Background \\ \hline
Total (after cuts) & 907 & 907 & 160 & 747 & 624 & 624 & 213 & 411 \\
BDT score $>$ 0 & 337 & 346 & 124 & 222 & 285 & 287 & 160 & 127 \\
BDT score $<$ 0 & 570 & 561 & 36 & 525 & 339 & 337 & 53 & 284 \\ \hline
\end{tabular}
\end{table*}

The differential cross section as a function of kinematic variable $X$ is defined as 
\begin{equation}\label{eq:sigma}
\frac{d\sigma}{dX} = \frac{N - N_{b}}{\Delta X N_{Ar}\Phi\epsilon_{\mathrm{UF}}}, 
\end{equation}
where $N$ is the number of data events in a given $X$ bin after applying the full selection, $N_{b}$ is the number of background event in the same bin, $\Delta X$ is the bin size, $N_{Ar}$ is the number of Ar nuclei inside the fiducial volume and $\Phi$ is the integrated flux during the data taking period. The bin-by-bin  unfolding factor $\epsilon_{\mathrm{UF}}$ combines corrections for acceptance, efficiencies of the event selection, and resolution effects using the CC1$\pi^{\pm}$ neutrino and antineutrino events simulated with GENIE. The numerator of the unfolding factor is obtained by applying the same requirements to the GENIE-simulated CC1$\pi^{\pm}$ events as the ones applied to data. The denominator of the unfolding factor is obtained by requiring the true neutrino interaction of the CC1$\pi^{\pm}$ event is inside the fiducial volume. In this paper, we report the differential cross sections as a function of four kinematic variables: the outgoing muon momentum ($p_{\mu}$), the outgoing muon angle with respect to the initial neutrino direction ($\theta_{\mu}$), the outgoing pion angle with respect to the initial neutrino direction ($\theta_{\pi}$) and the angle between the outgoing muon and pion ($\theta_{\mu \pi}$). 

\begin{figure*}[!ht]
\centering
\includegraphics[width=0.41\textwidth]{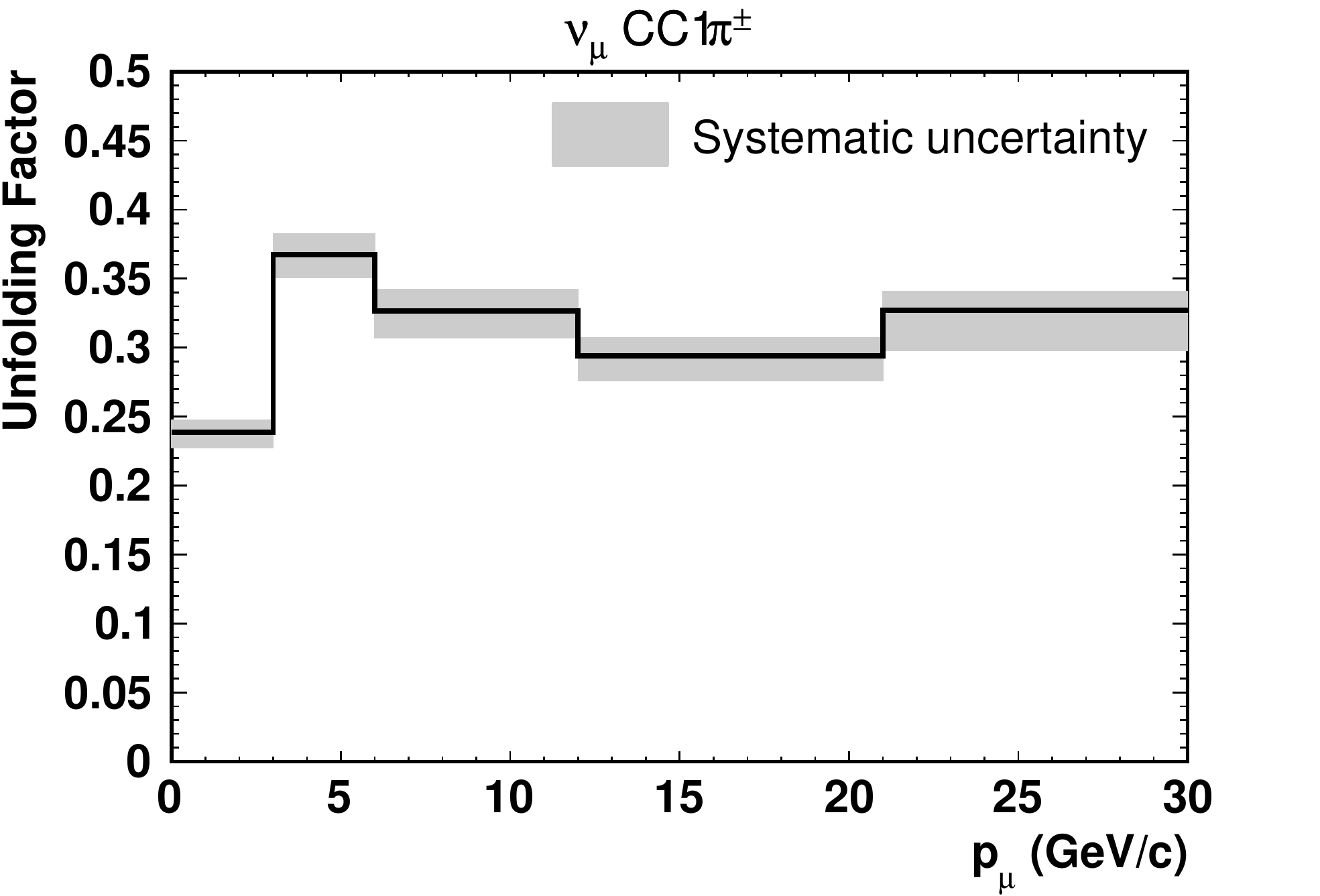}
\includegraphics[width=0.41\textwidth]{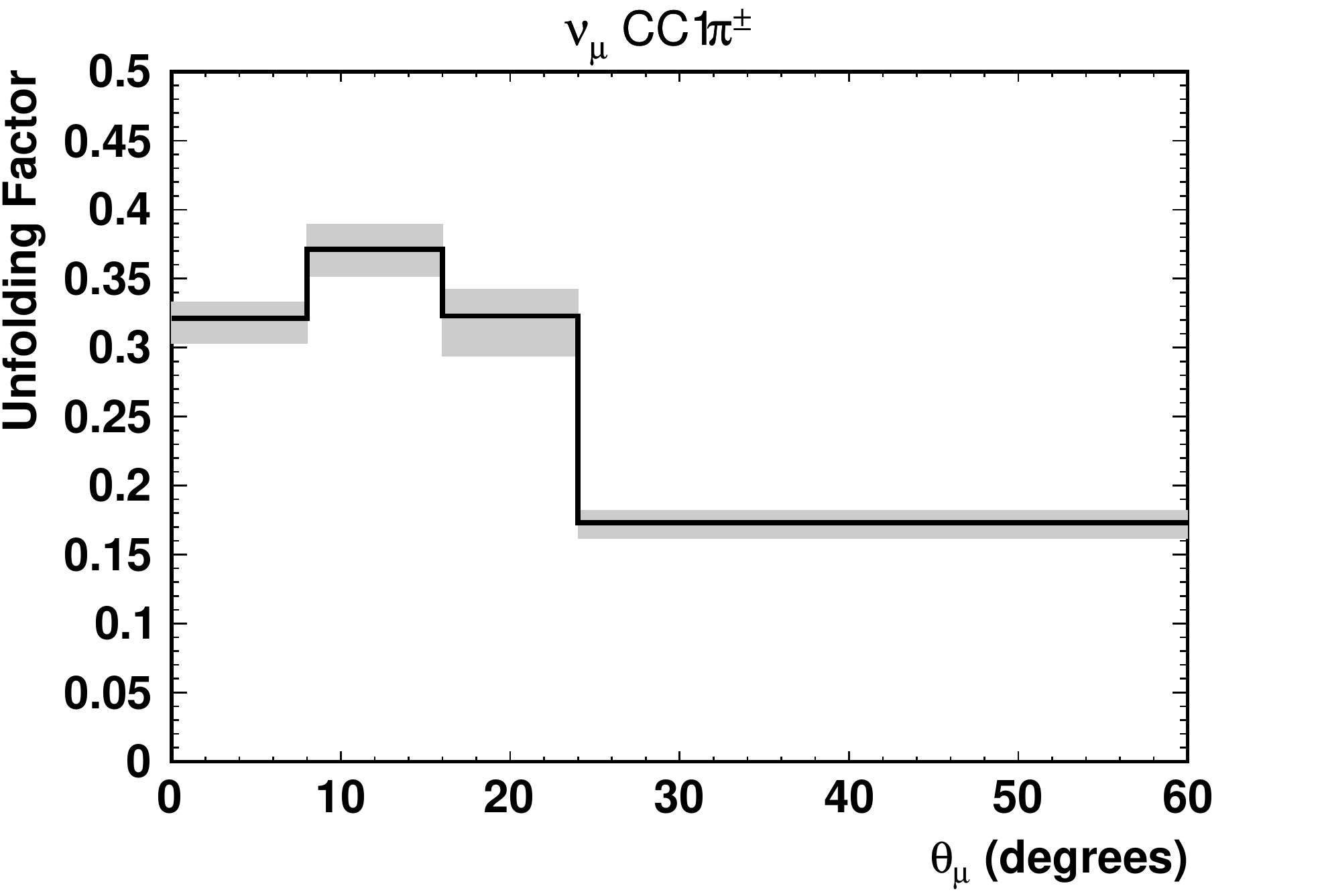}
\includegraphics[width=0.41\textwidth]{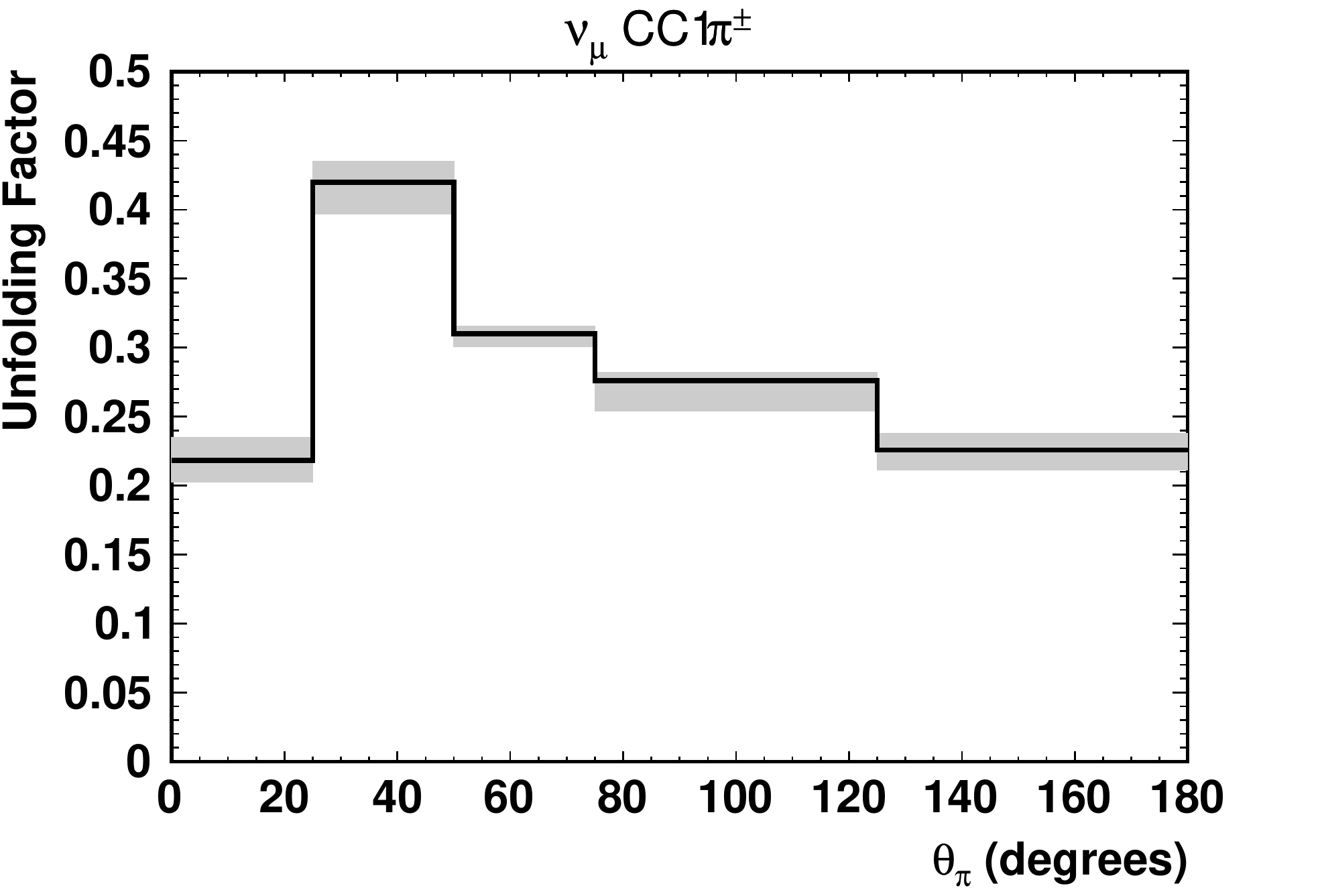}
\includegraphics[width=0.41\textwidth]{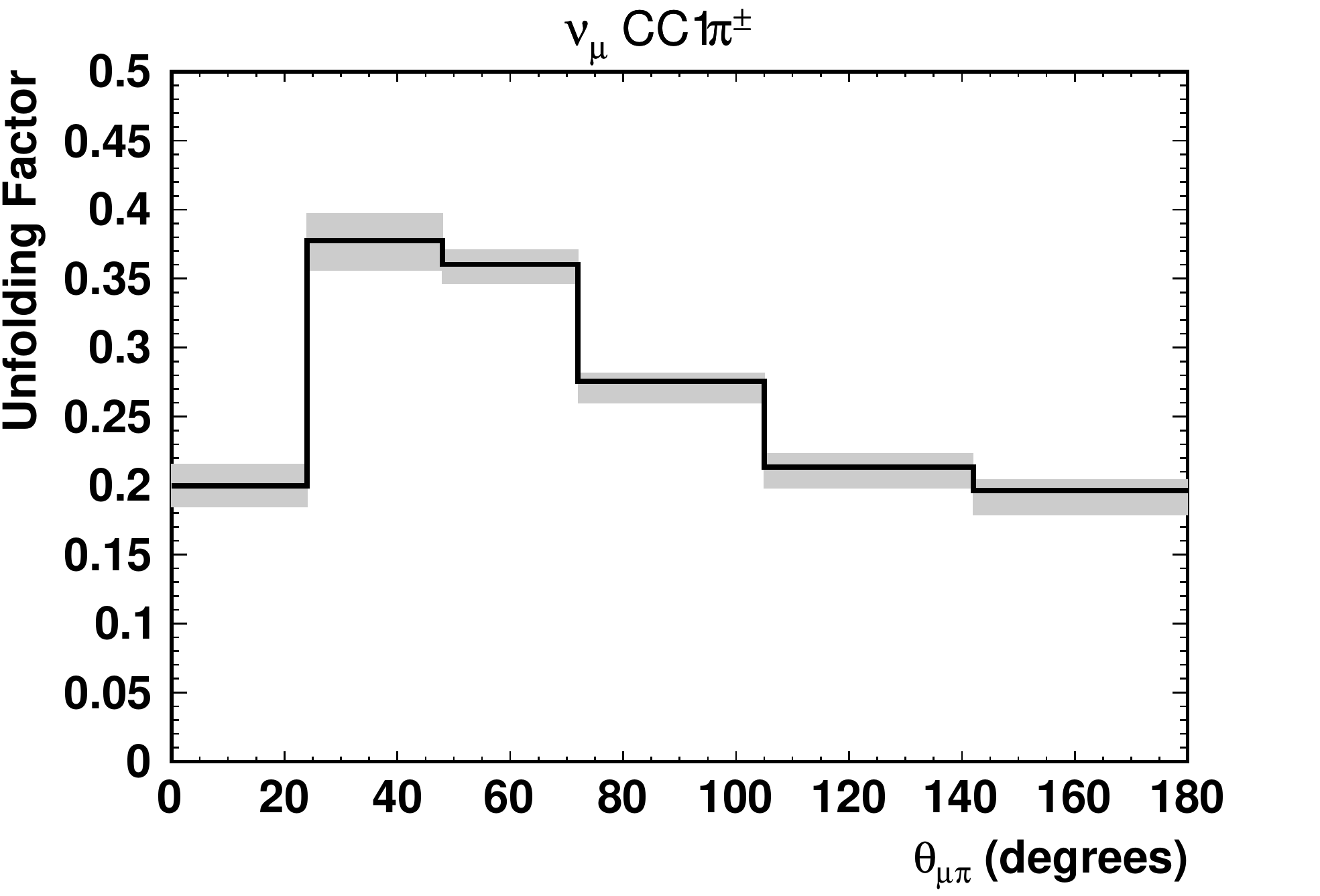}
\caption{Unfolding factors for $\nu_{\mu}$ CC1$\pi^{\pm}$ events. \commoncaption}\label{fig:eff_nu}
\end{figure*}

\begin{figure*}[!ht]
\centering
\includegraphics[width=0.41\textwidth]{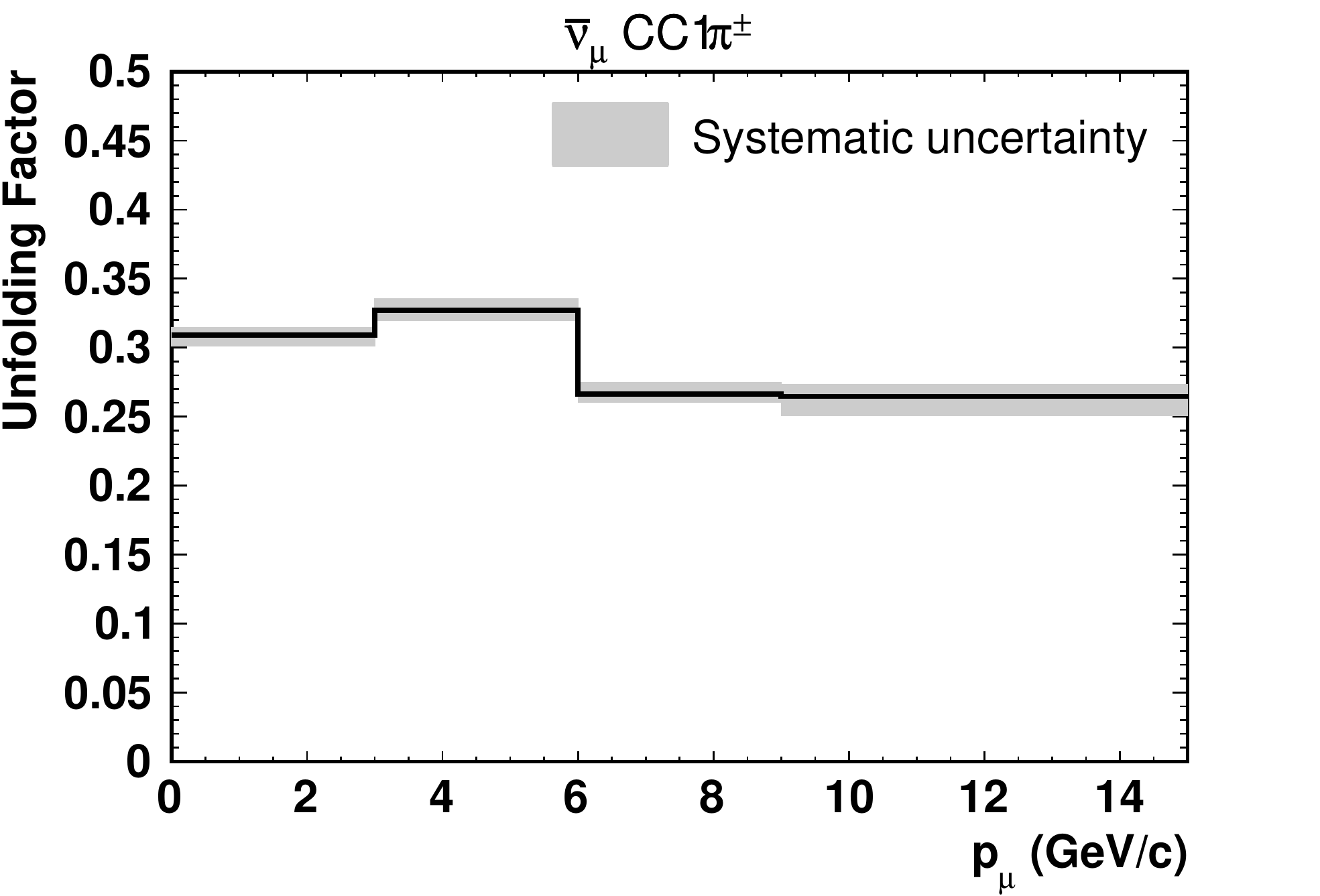}
\includegraphics[width=0.41\textwidth]{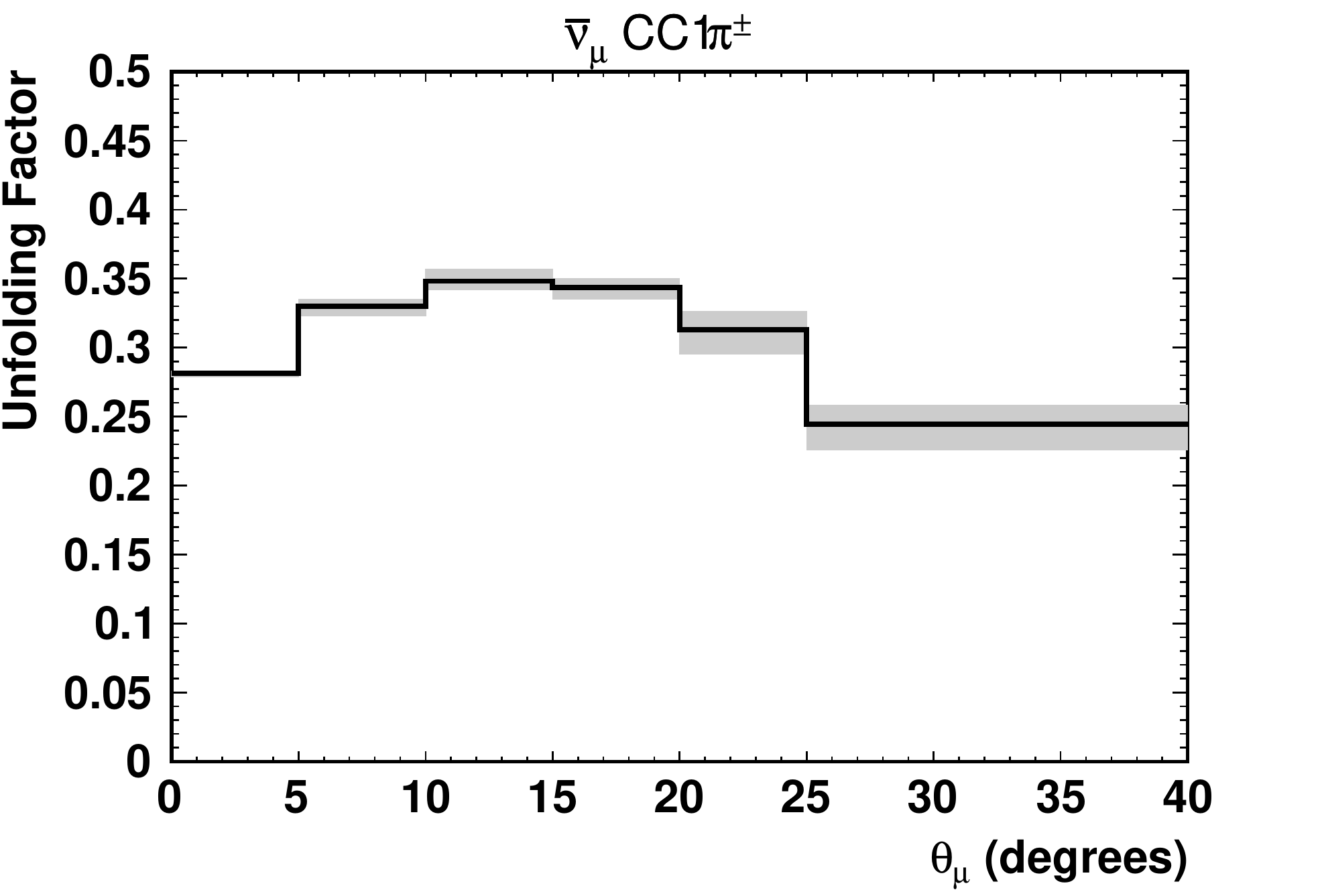}
\includegraphics[width=0.41\textwidth]{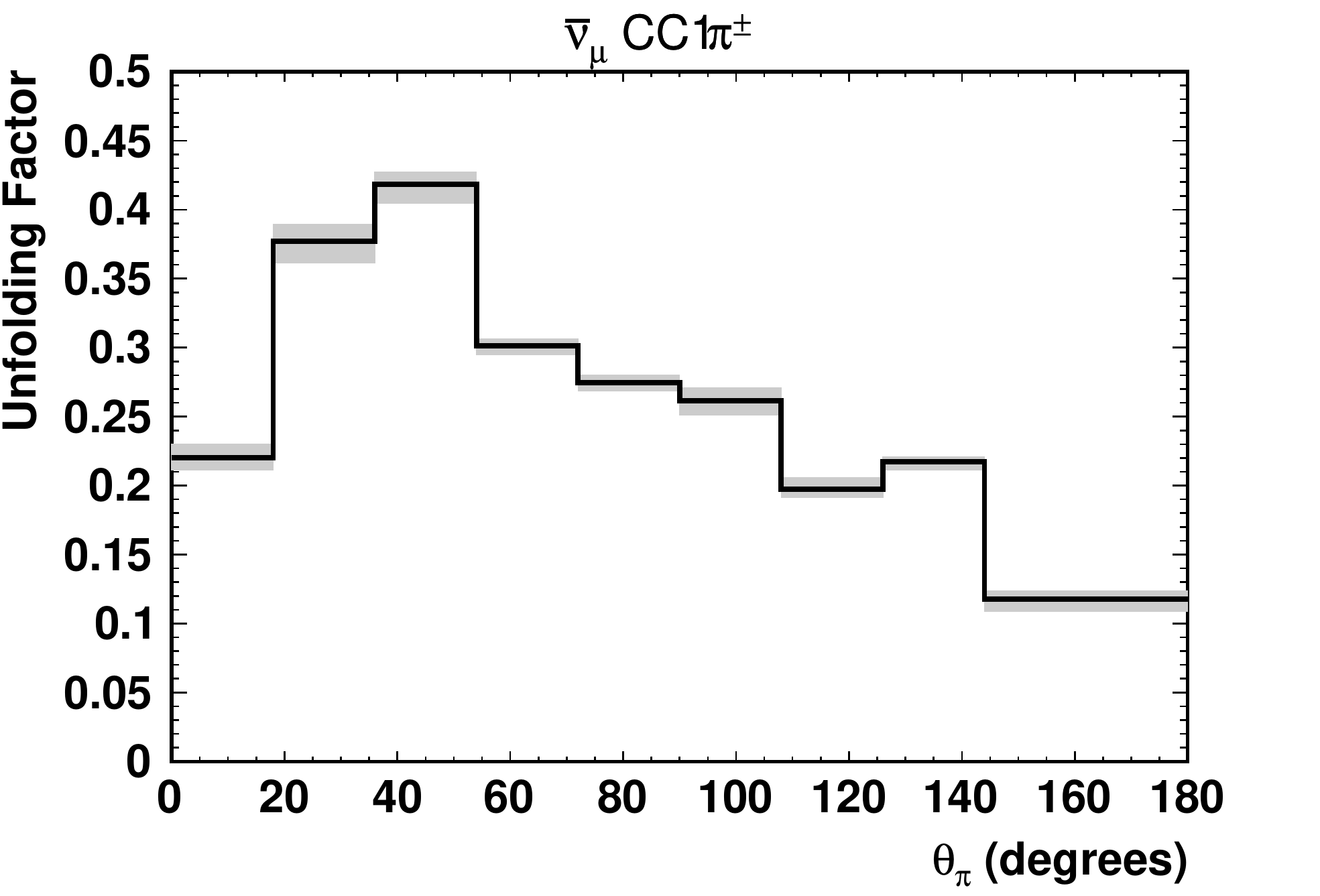}
\includegraphics[width=0.41\textwidth]{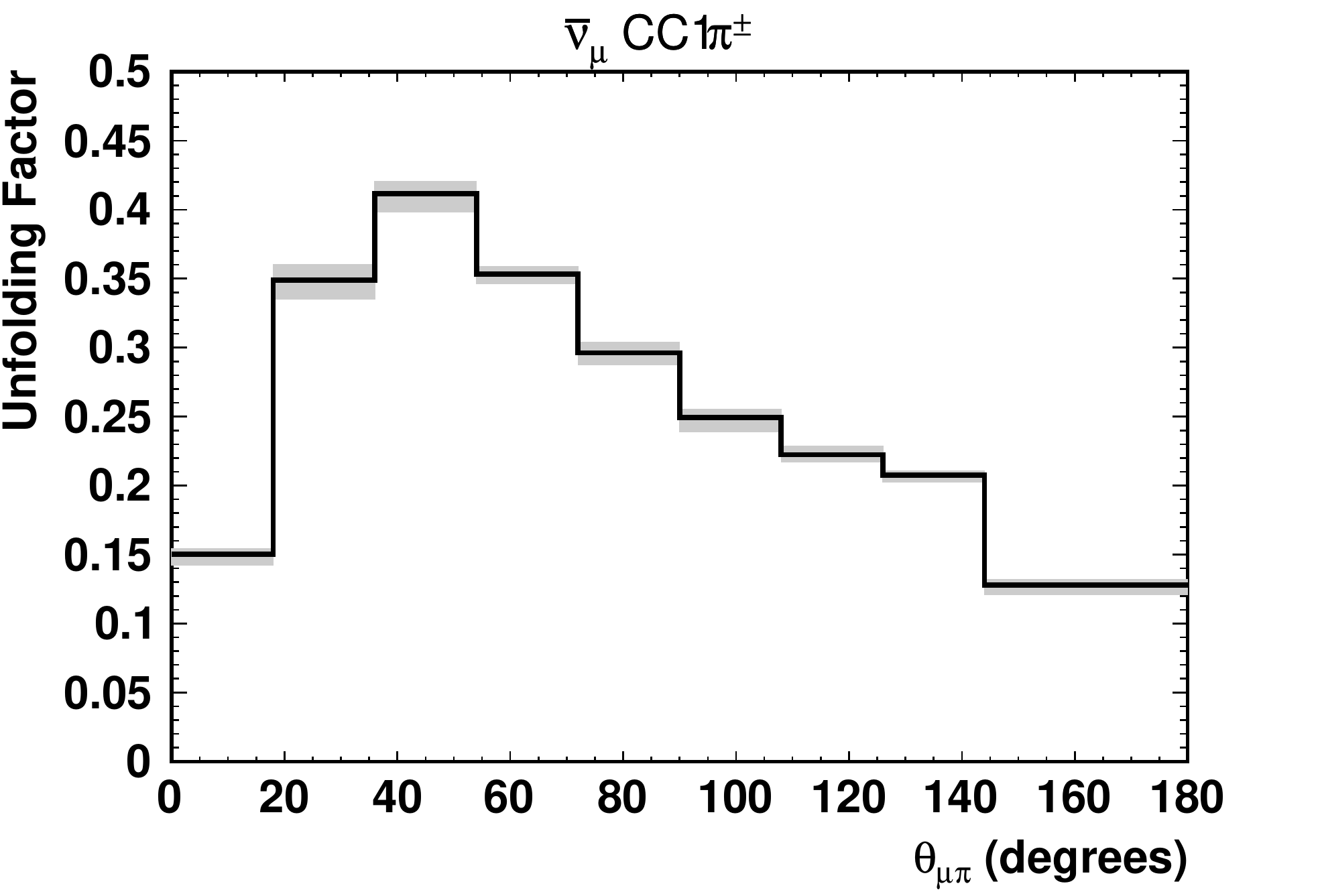}
\caption{Unfolding factors for $\bar{\nu}_{\mu}$ CC1$\pi^{\pm}$ events. \commoncaption}\label{fig:eff_anu}
\end{figure*}

Figures~\ref{fig:eff_nu} and \ref{fig:eff_anu} show the unfolding factors for $\nu_{\mu}$ and $\bar{\nu}_{\mu}$ CC1$\pi^{\pm}$ events, respectively. The bin sizes are choose to have reasonable statistics in each bin. The bin sizes are much bigger than the detector resolutions so the bin migration effect is negligible. The systematic uncertainties are evaluated by varying the GENIE simulation parameters, which will be discussed more later. The acceptance of muons goes from 80\% for forward going muons to 36\% for muon angle at 40 degrees.

\begin{figure*}[!ht]
\centering
\includegraphics[width=0.41\textwidth]{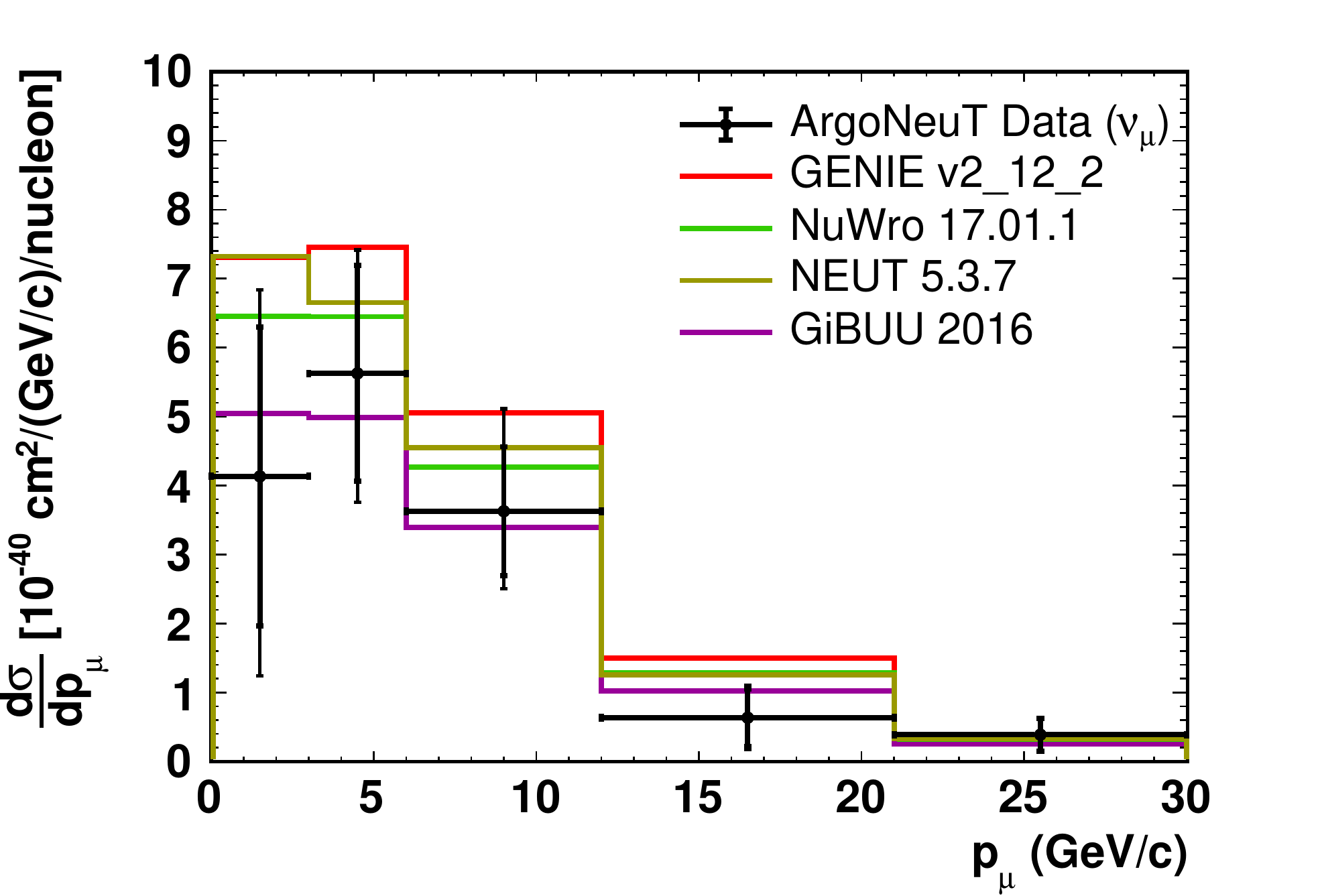}
\includegraphics[width=0.41\textwidth]{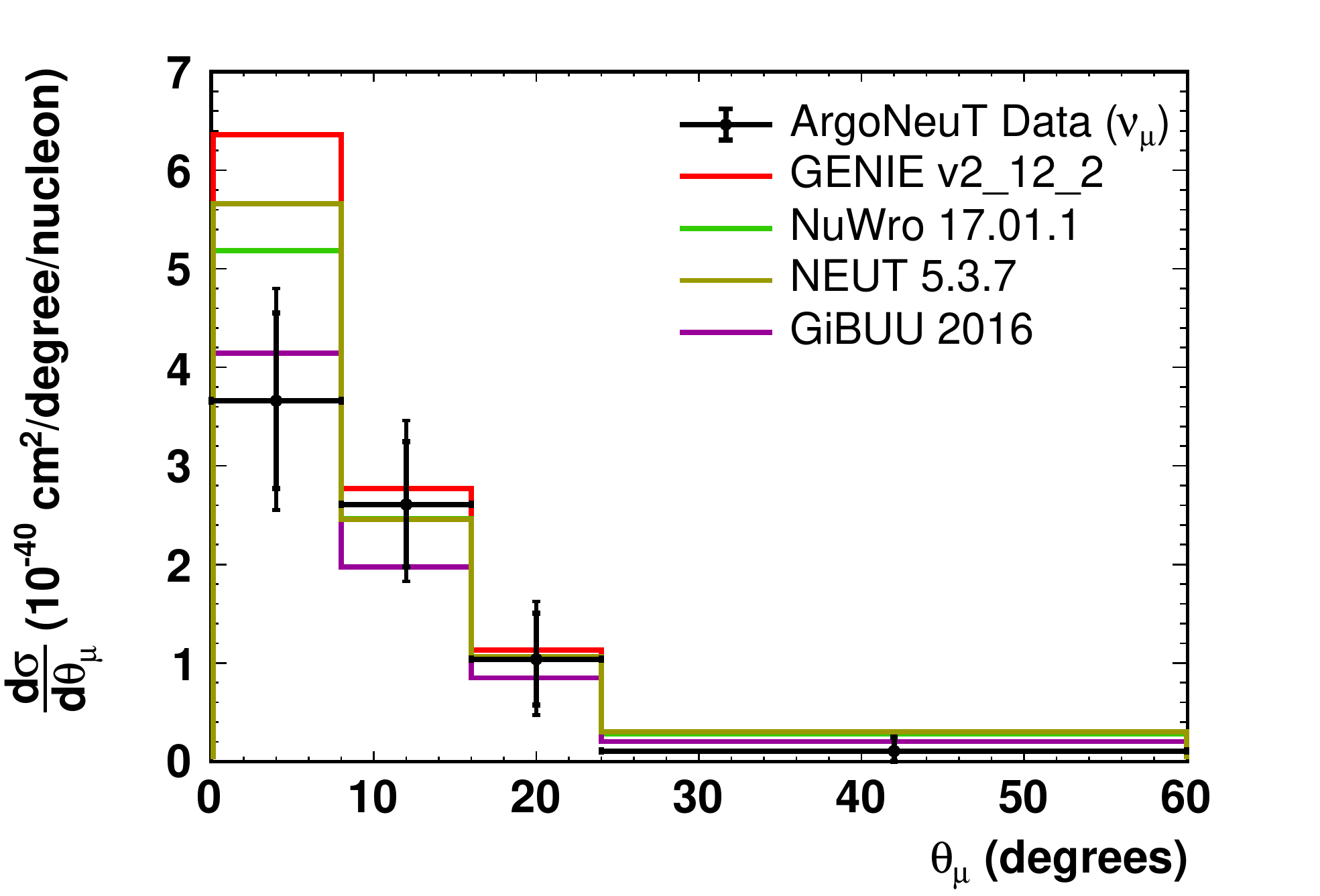}
\includegraphics[width=0.41\textwidth]{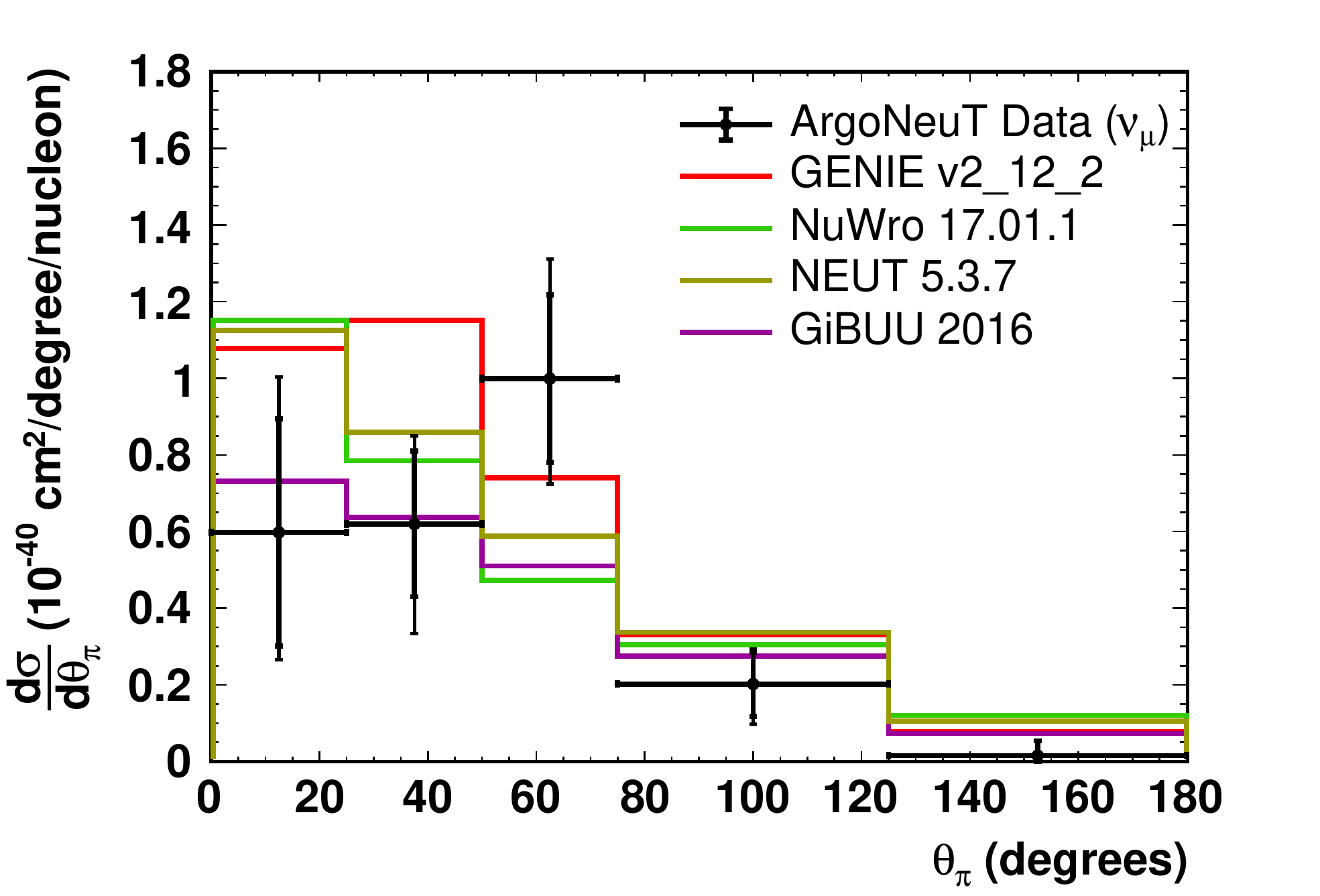}
\includegraphics[width=0.41\textwidth]{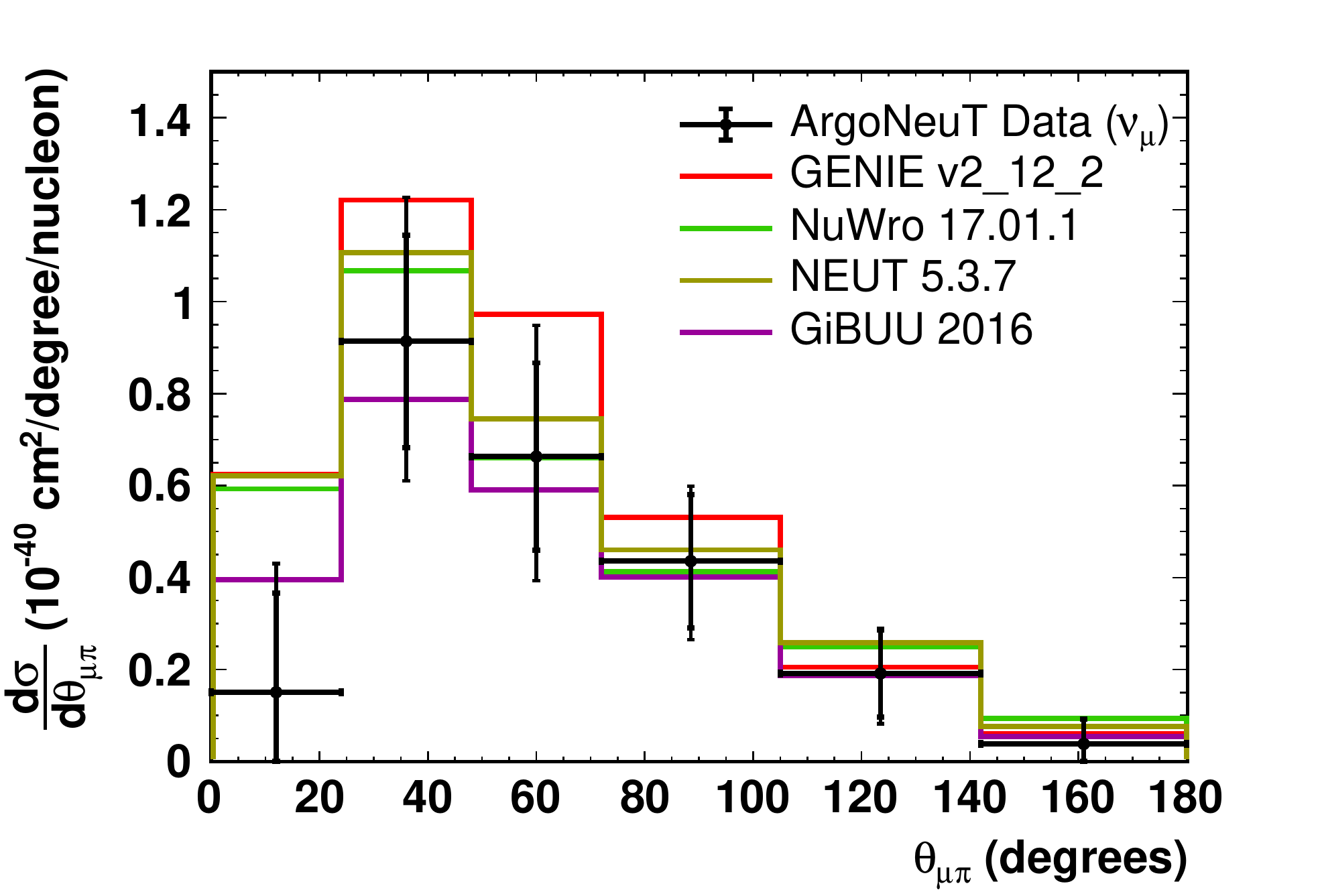}
\caption{ArgoNeuT $\nu_{\mu}$ CC1$\pi^{\pm}$ differential cross sections compared to GENIE, NuWro, GiBUU and NEUT. Thick error bars refers to statistical errors while thin error bars refers to statistical and systematic errors summed together. \commoncaption}\label{fig:xs_nu}
\end{figure*}

\begin{figure*}[!ht]
\centering
\includegraphics[width=0.41\textwidth]{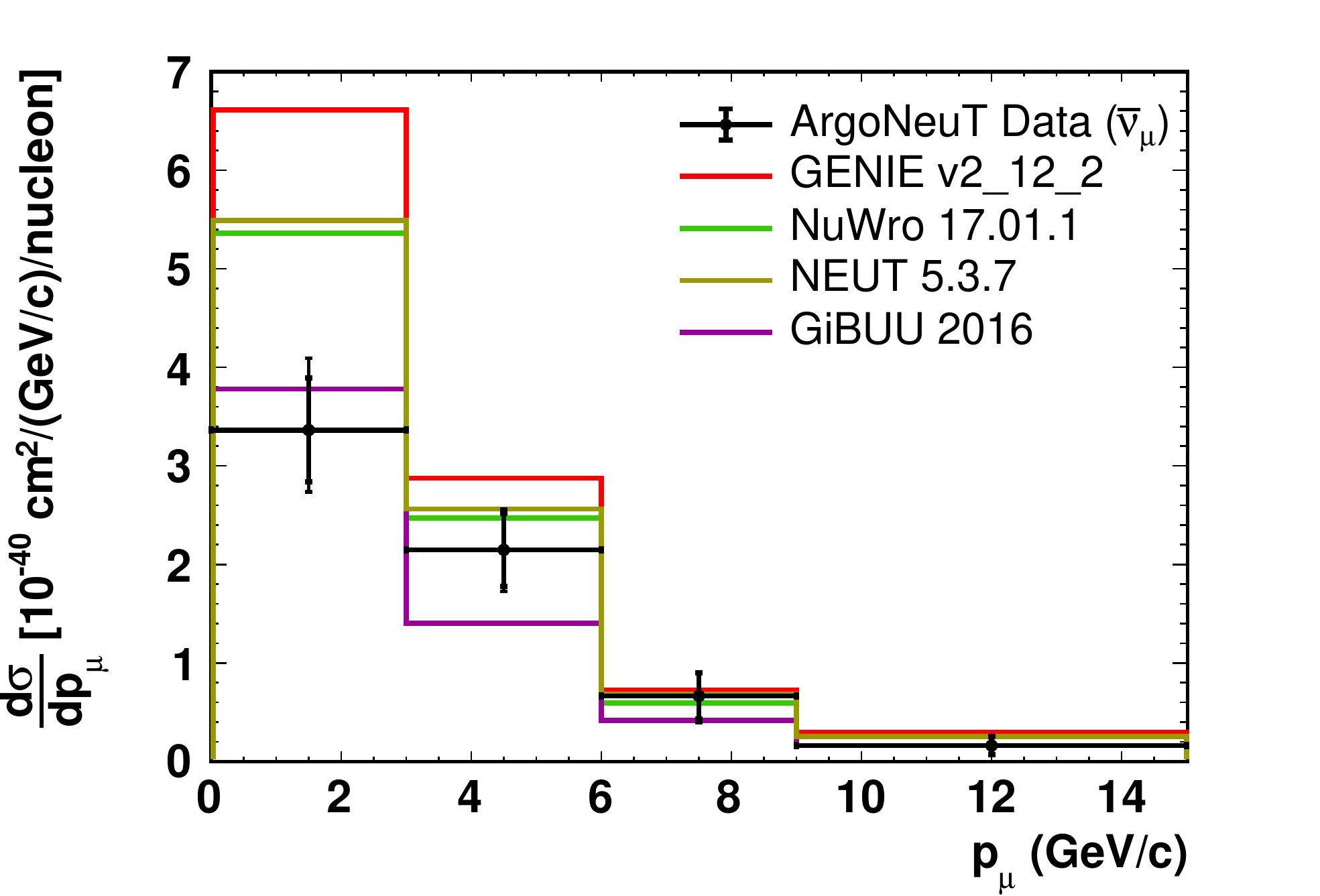}
\includegraphics[width=0.41\textwidth]{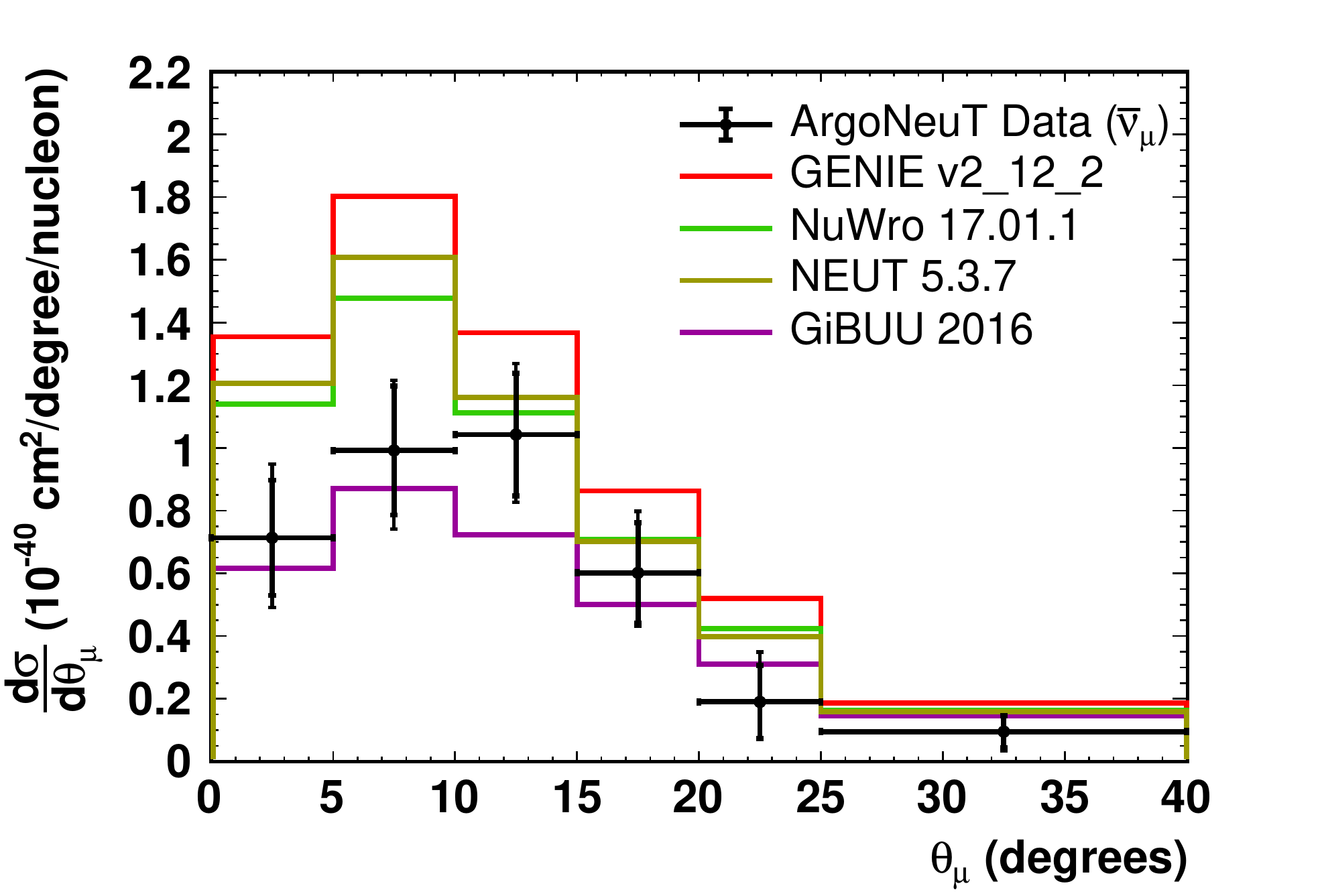}
\includegraphics[width=0.41\textwidth]{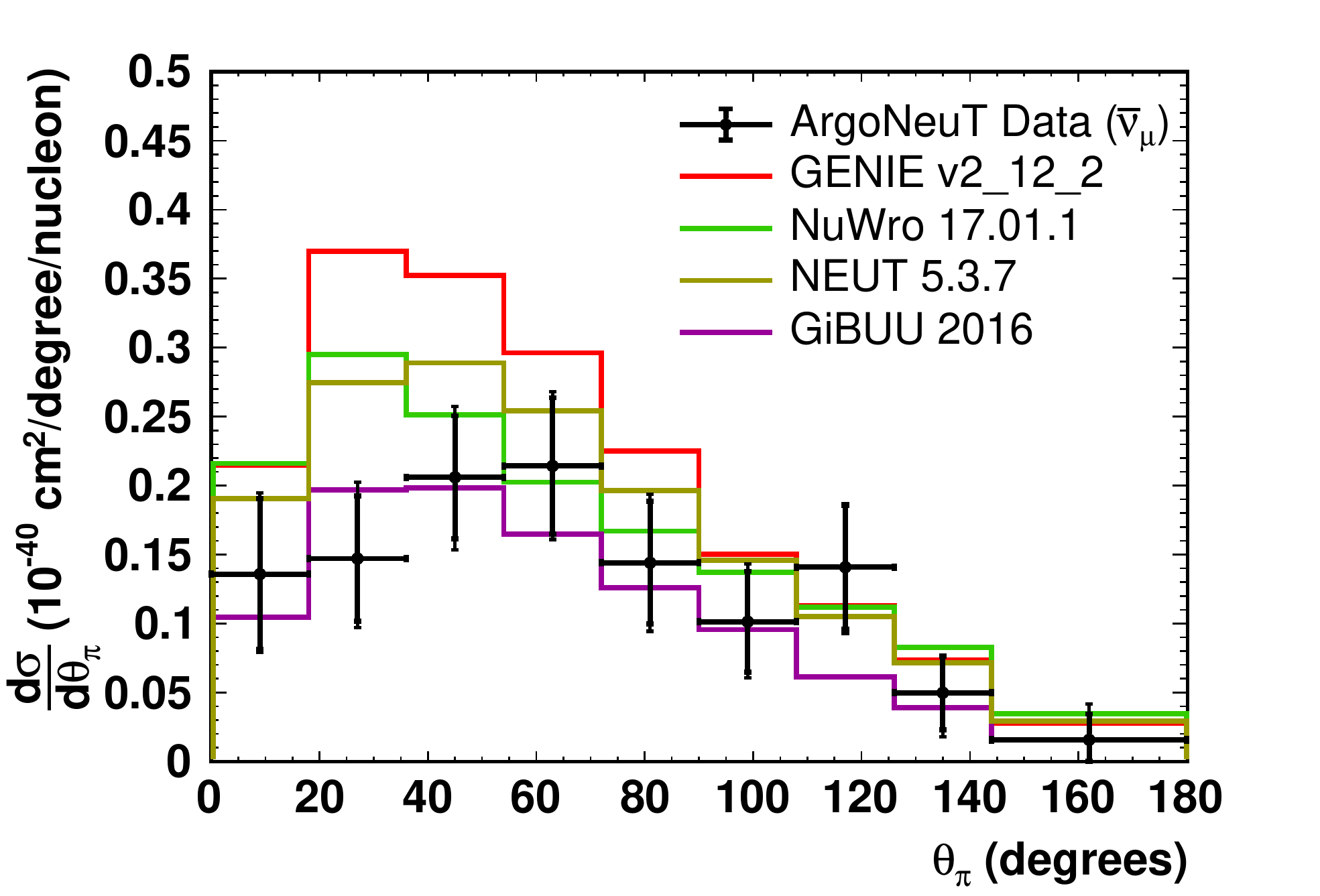}
\includegraphics[width=0.41\textwidth]{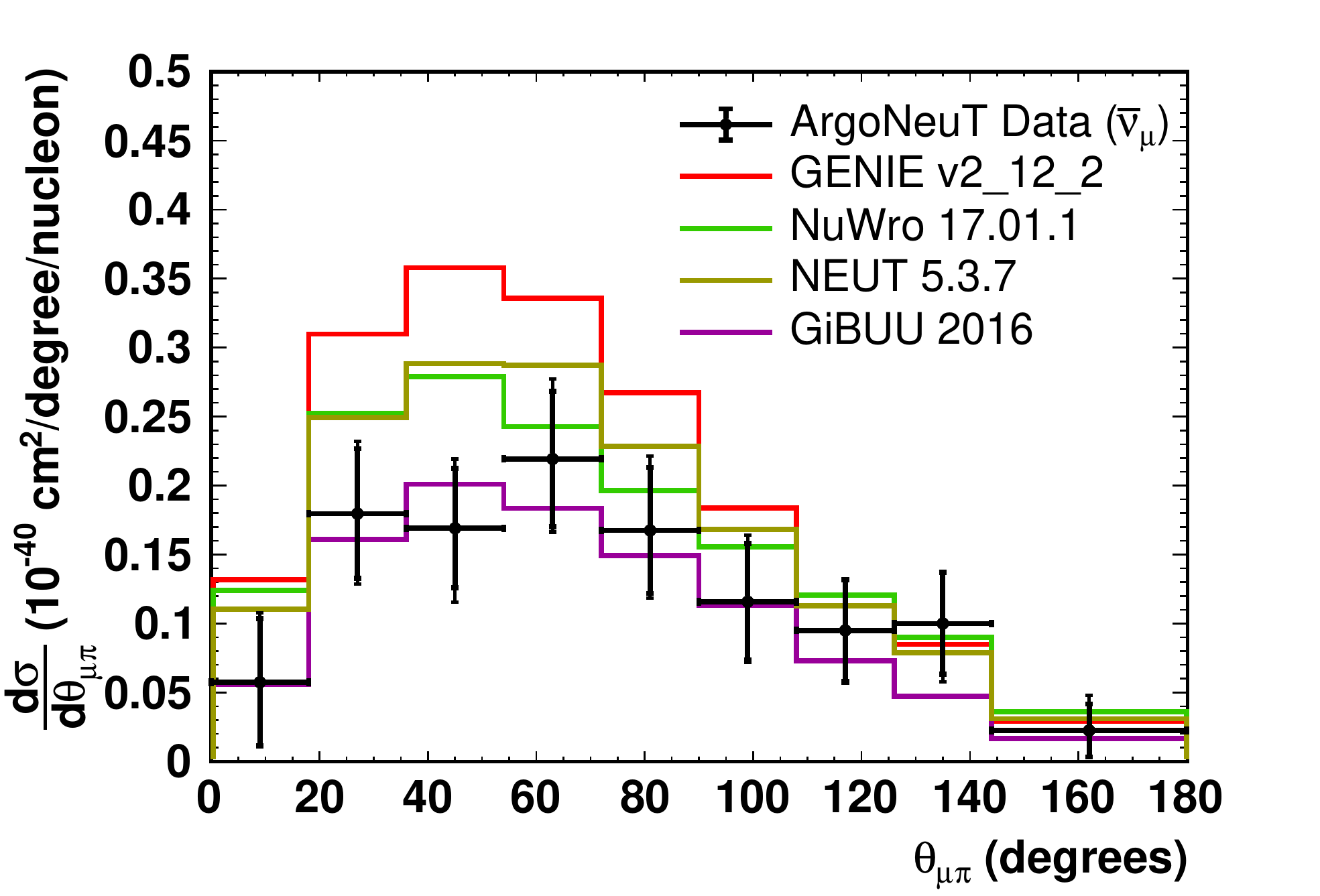}
\caption{ArgoNeuT $\bar{\nu}_{\mu}$ CC1$\pi^{\pm}$ differential cross sections compared to GENIE, NuWro, GiBUU and NEUT. Thick error bars refers to statistical errors while thin error bars refers to statistical and systematic errors summed together. \commoncaption}\label{fig:xs_anu} 
\end{figure*}

The measured differential cross sections for $\nu_{\mu}$($\bar{\nu}_{\mu}$) CC1$\pi^{\pm}$ productions are shown in Fig.~\ref{fig:xs_nu}(\ref{fig:xs_anu}) and listed in Table~\ref{tab:xs_nu}(\ref{tab:xs_anu}). The flux-averaged cross sections are listed in Table~\ref{tab:xs}. Both the differential and flux-averaged cross sections are compared with four MC generators: GENIE, GiBUU~\cite{gibuu}, NEUT~\cite{Hayato:2009zz, Abe:2017aap}, and NuWro~\cite{Golan:2012wx}. Predictions from the different MC generators are shown in Fig.~\ref{fig:xs_nu} and~\ref{fig:xs_anu}. 

\begin{table*}[!ht]
\centering
\begin{minipage}[t]{\columnwidth}
\centering
  \caption{The measured differential cross sections in $p_{\mu}$, $\theta_{\mu}$, $\theta_{\pi}$ and $\theta_{\mu\pi}$ for $\nu_{\mu}$ CC1$\pi^{\pm}$ interactions in argon. Both statistical (first) and systematic (second) errors are shown.}\label{tab:xs_nu}
%  \begin{subfigure}[b]{0.5\textwidth}
\begin{tabular}{@{}cc@{}}
\hline
$p_{\mu}$ & $d\sigma/dp_{\mu}$ ($\nu_{\mu}$) \\ 
{[}GeV/c{]} & {[}10$^{-37}$ cm$^{2}$/(GeV/c)/nucleon{]} \\ 
\hline
0-3 & 4.13 $\pm$ 2.17 $^{+1.62}_{-1.91}$ \\ 
3-6 & 5.63 $\pm$ 1.56 $^{+0.87}_{-1.02}$ \\ 
6-12 & 3.63 $\pm$ 0.94 $^{+1.15}_{-0.63}$ \\ 
12-21 & 0.63 $\pm$ 0.42 $^{+0.21}_{-0.20}$ \\ 
21-30 & 0.39 $\pm$ 0.24 $^{+0.06}_{-0.06}$ \\ 
\hline
\end{tabular}
%  \end{subfigure}
%  \hspace{3mm}
%  \begin{subfigure}[b]{0.5\textwidth}
\begin{tabular}{@{}cc@{}}
\hline
$\theta_{\mu}$   & $d\sigma/d\theta_{\mu}$ ($\nu_{\mu}$) \\ 
{[}degrees{]} & {[}10$^{-37}$ cm$^{2}$/degree/nucleon{]} \\ 
\hline
0-8 & 3.66 $\pm$ 0.89 $^{+0.71}_{-0.66}$ \\ 
8-16 & 2.61 $\pm$ 0.64 $^{+0.57}_{-0.45}$ \\ 
16-24 & 1.04 $\pm$ 0.46 $^{+0.36}_{-0.33}$ \\ 
24-60 & 0.11 $\pm$ 0.14 $^{+0.05}_{-0.11}$ \\ 
\hline
\end{tabular}   
%  \end{subfigure}
%  \vspace{7mm}
%  \begin{subfigure}[b]{0.5\textwidth}
\begin{tabular}{@{}cc@{}}
\hline
$\theta_{\pi}$   & $d\sigma/d\theta_{\pi}$ ($\nu_{\mu}$) \\ 
{[}degrees{]} & {[}10$^{-37}$ cm$^{2}$/degree/nucleon{]} \\ 
\hline
0-25 & 0.60 $\pm$ 0.30 $^{+0.28}_{-0.15}$ \\ 
25-50 & 0.62 $\pm$ 0.19 $^{+0.13}_{-0.21}$ \\ 
50-75 & 1.01 $\pm$ 0.22 $^{+0.22}_{-0.17}$ \\ 
75-125 & 0.20 $\pm$ 0.08 $^{+0.04}_{-0.06}$ \\ 
125-180 & 0.02 $\pm$ 0.04 $^{+0.01}_{-0.01}$ \\ 
\hline
\end{tabular}   
%  \end{subfigure}
%  \hspace{3mm}
%  \begin{subfigure}[b]{0.5\textwidth}
\begin{tabular}{@{}cc@{}}
\hline
$\theta_{\mu\pi}$   & $d\sigma/d\theta_{\mu\pi}$ ($\nu_{\mu}$) \\ 
{[}degrees{]} & {[}10$^{-37}$ cm$^{2}$/degree/nucleon{]} \\ 
\hline
0-24 & 0.15 $\pm$ 0.22 $^{+0.18}_{-0.06}$ \\ 
24-48 & 0.91 $\pm$ 0.23 $^{+0.21}_{-0.20}$ \\ 
48-72 & 0.66 $\pm$ 0.20 $^{+0.20}_{-0.18}$ \\ 
72-105 & 0.44 $\pm$ 0.14 $^{+0.07}_{-0.09}$ \\ 
105-142 & 0.19 $\pm$ 0.09 $^{+0.02}_{-0.06}$ \\ 
142-180 & 0.04 $\pm$ 0.05 $^{+0.01}_{-0.01}$ \\ 
\hline
\end{tabular}
\end{minipage}\hfill
%  \end{subfigure}
\begin{minipage}[t]{\columnwidth}
\caption{The measured differential cross sections in $p_{\mu}$, $\theta_{\mu}$, $\theta_{\pi}$ and $\theta_{\mu\pi}$ for $\bar{\nu}_{\mu}$ CC1$\pi^{\pm}$ interactions in argon. Both statistical (first) and systematic (second) errors are shown.}\label{tab:xs_anu}
\centering
%  \begin{subfigure}[b]{0.5\textwidth}
\begin{tabular}{@{}cc@{}}
\hline
$p_{\mu}$ & $d\sigma/dp_{\mu}$ ($\overline{\nu}_{\mu}$) \\ 
{[}GeV/c{]} & {[}10$^{-38}$ cm$^{2}$/(GeV/c)/nucleon{]} \\ 
\hline
0-3 & 3.36 $\pm$ 0.53 $^{+0.50}_{-0.34}$ \\ 
3-6 & 2.15 $\pm$ 0.37 $^{+0.19}_{-0.20}$ \\ 
6-9 & 0.66 $\pm$ 0.23 $^{+0.05}_{-0.14}$ \\ 
9-15 & 0.16 $\pm$ 0.09 $^{+0.02}_{-0.02}$ \\ 
\hline
\end{tabular}
%  \end{subfigure}
%  \hspace{3mm}
%  \begin{subfigure}[b]{0.5\textwidth}
\begin{tabular}{@{}cc@{}}
\hline
$\theta_{\mu}$ & $d\sigma/d\theta_{\mu}$ ($\overline{\nu}_{\mu}$) \\ 
{[}degrees{]} & {[}10$^{-38}$ cm$^{2}$/degree/nucleon{]} \\ 
0-5 & 0.71 $\pm$ 0.18 $^{+0.15}_{-0.12}$ \\ 
5-10 & 1.00 $\pm$ 0.21 $^{+0.09}_{-0.14}$ \\ 
10-15 & 1.04 $\pm$ 0.20 $^{+0.11}_{-0.09}$ \\ 
15-20 & 0.60 $\pm$ 0.16 $^{+0.11}_{-0.06}$ \\ 
20-25 & 0.19 $\pm$ 0.12 $^{+0.11}_{-0.03}$ \\ 
25-40 & 0.10 $\pm$ 0.05 $^{+0.01}_{-0.03}$ \\ 
\hline
\end{tabular}
%  \end{subfigure}
%  \vspace{7mm}
%  \begin{subfigure}[b]{0.5\textwidth}
\begin{tabular}{@{}cc@{}}
\hline
$\theta_{\pi}$ & $d\sigma/d\theta_{\pi}$ ($\overline{\nu}_{\mu}$) \\ 
{[}degrees{]} & {[}10$^{-38}$ cm$^{2}$/degree/nucleon{]} \\ 
\hline
0-18 & 0.14 $\pm$ 0.05 $^{+0.02}_{-0.02}$ \\ 
18-36 & 0.15 $\pm$ 0.05 $^{+0.03}_{-0.02}$ \\ 
36-54 & 0.21 $\pm$ 0.04 $^{+0.03}_{-0.03}$ \\ 
54-72 & 0.21 $\pm$ 0.05 $^{+0.02}_{-0.02}$ \\ 
72-90 & 0.14 $\pm$ 0.04 $^{+0.02}_{-0.02}$ \\ 
90-108 & 0.10 $\pm$ 0.04 $^{+0.02}_{-0.02}$ \\ 
108-126 & 0.14 $\pm$ 0.04 $^{+0.01}_{-0.02}$ \\ 
126-144 & 0.05 $\pm$ 0.03 $^{+0.01}_{-0.02}$ \\ 
144-180 & 0.02 $\pm$ 0.02 $^{+0.02}_{-0.01}$ \\ 
\hline
\end{tabular}
%  \end{subfigure}
%  \hspace{3mm}
%  \begin{subfigure}[b]{0.5\textwidth}
\begin{tabular}{@{}cc@{}}
\hline
$\theta_{\mu\pi}$ & $d\sigma/d\theta_{\mu\pi}$ ($\overline{\nu}_{\mu}$) \\ 
{[}degrees{]} & {[}10$^{-38}$ cm$^{2}$/degree/nucleon{]} \\ 
\hline
0-18 & 0.06 $\pm$ 0.05 $^{+0.02}_{-0.01}$ \\ 
18-36 & 0.18 $\pm$ 0.05 $^{+0.02}_{-0.02}$ \\ 
36-54 & 0.17 $\pm$ 0.04 $^{+0.03}_{-0.03}$ \\ 
54-72 & 0.22 $\pm$ 0.05 $^{+0.03}_{-0.02}$ \\ 
72-90 & 0.17 $\pm$ 0.05 $^{+0.03}_{-0.02}$ \\ 
90-108 & 0.12 $\pm$ 0.04 $^{+0.02}_{-0.01}$ \\ 
108-126 & 0.09 $\pm$ 0.04 $^{+0.01}_{-0.01}$ \\ 
126-144 & 0.10 $\pm$ 0.04 $^{+0.01}_{-0.02}$ \\ 
144-180 & 0.02 $\pm$ 0.02 $^{+0.02}_{-0.01}$ \\ 
\hline
\end{tabular}
\end{minipage}
%\end{subfigure}
\end{table*}

\begin{table*}[!ht]
\centering
\caption{Comparison between measured flux-averaged cross sections and MC generator predictions for neutrinos at a mean energy of 9.6 GeV and antineutrinos at a mean energy of 3.6 GeV.}\label{tab:xs}
\begin{tabular}{c c c}
\hline
&\multicolumn{2}{c}{Flux-averaged CC1$\pi$ cross section} \\ 
\hline
 & $\nu_{\mu}$ [10$^{-37}$ cm$^{2}$/Ar] & $\bar{\nu}_{\mu}$ [10$^{-38}$ cm$^{2}$/Ar] \\ 
ArgoNeuT data & 2.7 $\pm$ 0.5 (stat) $\pm$ 0.5 (syst) & 8.4 $\pm$ 0.9 (stat) $_{-0.8}^{+1.0}$ (syst) \\
\hline
GENIE & 3.8 & 13.3 \\
NuWro & 3.3 & 11.0 \\
GiBUU & 2.6 &7.3 \\
NEUT & 3.5 & 11.4 \\ 
\hline
\end{tabular}
\end{table*}  

The systematic uncertainties affecting the background estimation, unfolding and final cross section measurements are listed in Table~\ref{tab:syst}. These are dominated by the neutrino flux-scale uncertainty and GENIE modeling uncertainties. The flux normalization uncertainties are taken to be 9.7\% and 7.8\% respectively for neutrino and antineutrino~\cite{Aliaga:2016oaz}. The GENIE modeling uncertainties were obtained by varying many parameters according to their uncertainties in the generator. The effects we consider include the uncertainties in quasielastic scattering, resonant pion production, coherent pion production, overlap between resonant pion production and deep inelastic scattering, and FSI processes. The larger background contamination and more nonresonant contribution to the signal events in the neutrino sample make it more susceptible to the GENIE modeling uncertainties. The energy scale uncertainty is taken to be 3\% based on the calorimetric reconstruction of the through-going muons, see Fig.~\ref{fig:dedxmu}. The uncertainty on the number of argon targets is taken as 2\%, which corresponds to a 4 mm uncertainty on each dimension of the fiducial volume definition. The uncertainty on POT is taken as 1\%~\cite{Anderson:2011ce, Acciarri:2014isz}. 

\begin{table*}[!ht]
\centering
\caption{List of systematic errors affecting this analysis and their estimation.}\label{tab:syst}
\begin{tabular}{c c c}
\hline
 & \multicolumn{2}{c}{Cross section uncertainty (\%)} \\ 
 \hline
Systematic Uncertainty & $\nu_{\mu}$ & $\overline{\nu}_{\mu}$ \\ 
\hline
Flux Normalization & $^{+11.0}_{-9.0}$ & $^{+8.0}_{-7.0}$ \\ 
GENIE modeling & $^{+14.5}_{-16.6}$& $^{+5.5}_{-5.2}$\\
Energy scale & $^{+7.7}_{-0.0}$ & $^{+7.2}_{-1.8}$\\ 
Number of argon targets & $\pm2.0$ & $\pm2.0$ \\ 
POT & $\pm1.0$ & $\pm1.0$ \\ 
\hline
Total systematics & $^{+19.9}_{-19.0}$ & $^{+12.3}_{-9.2}$\\ 
\hline
\end{tabular}
\end{table*}

Each generator includes models for the initial neutrino interaction, the nuclear structure affecting the initial interaction, two-particle two-hole (2p2h) excitation and the FSI of the particles produced. GENIE and NEUT use the model of Rein and Sehgal~\cite{Rein:1980wg} to describe pion production. NuWro includes only the $\Delta$(1232) resonance according to the model~\cite{Jones:1972ky}. NEUT takes the nonresonant interaction from Rein and Sehgal; GENIE and NuWro use the model of Bodek and Yang~\cite{Bodek:2004pc} above the resonance region and smoothly extrapolate it to lower hadronic invariant mass ($W$) to converge with the resonance model. For FSI, NEUT and NuWro use the Salcedo-Oset model~\cite{Salcedo:1987md} in a cascade formalism which has nuclear medium corrections, while GENIE uses an effective cascade model. GiBUU uses quantum-kinetic transport theory, which allows one to include important nuclear effects such as binding potentials for hadrons and spectral functions, including their dynamical development. GiBUU also requires consistency in the sense that the description of all subprocesses, such as, e.g., quasielastic scattering, pion production, deep inelastic scattering, and 2p2h interactions, is based on the same ground state. The modeling of pion production in GiBUU is described in detail in~\cite{Lalakulich:2012cj, Lalakulich:2010ss}. 

The measured neutrino cross sections have larger statistical and systematic uncertainties compared with the measured antineutrino cross sections because of the larger background contamination in the neutrino sample. According to Table~\ref{tab:bdtfit}, the BDT score $>$ 0 region is dominated by background for the $\nu_{\mu}$ events. Since we rely on MC simulations to model the background shape in the BDT fit, the result is more susceptible to the GENIE modeling uncertainties that affect various background contribution. 

The GiBUU predictions are in good agreement with the measured cross sections. The NuWro and NEUT predictions are similar to each other and both higher than the measured cross section presented here. The GENIE predictions appear to be higher than the other generator predictions and the measured cross sections. This is consistent with the conclusion in~\cite{Rodrigues:2016xjj} that GENIE's non-resonant background prediction has to be significantly reduced to fit the data. We do not change any default GENIE parameters in this analysis. All predicted integrated cross sections agree with the measurements within 2 $\sigma$ of the total uncertainty except the GENIE $\bar{\nu}_{\mu}$ CC1$\pi^{\pm}$ prediction which deviates from the measurement by 3.3 $\sigma$. In general all the predictions agree better with the measured neutrino cross sections than the measured antineutrino cross sections. This is expected because there have been a few measurements of the neutrino CC$1\pi^{\pm}$ cross sections on nuclear targets but no antineutrino measurements. Therefore the generators could not have been tuned on antineutrino data as they are on neutrino data.  This makes the measured antineutrino CC$1\pi^{\pm}$ cross sections reported in this paper valuable to the improvements of MC generators.

\section{\label{sec:summary}Summary}
This paper presents the measurements of neutrino- and antineutrino-induced pion production on an argon target and compares them to different MC generators. The flux-averaged cross sections are measured to be
$
2.7\pm0.5(stat)\pm0.5(syst) \times 10^{-37} \textrm{cm}^{2}/\textrm{Ar}
$
for neutrinos at a mean energy of 9.6 GeV and
$ 
8.4\pm0.9(stat)^{+1.0}_{-0.8}(syst) \times 10^{-38} \textrm{cm}^{2}/\textrm{Ar}
$
for antineutrinos at a mean energy of 3.6 GeV with the charged pion momentum above 100 MeV/$c$. These measurements provide new information about the neutrino single pion production and can be used to improve the modeling of neutrino interactions with the argon nucleus.

\section*{Acknowledgement}
This manuscript has been authored by Fermi Research Alliance, LLC under Contract No. DE-AC02-07CH11359 with the U.S. Department of Energy, Office of Science, Office of High Energy Physics.
We are grateful to C.~Andreopoulos, Y.~Hayato, U.~Mosel and J.~Sobczyk for providing theoretical predictions and
for many useful discussions. We gratefully acknowledge the cooperation of the MINOS Collaboration in providing their data for use in this analysis. We wish to acknowledge the support of Fermilab, the Department of Energy, and the National Science Foundation in ArgoNeuT's construction, operation, and data analysis. We also wish to acknowledge the support of the Neutrino Physics Center (NPC) Scholar program at Fermilab.

\clearpage

\appendix

\section{\label{sec:reco}Event Reconstruction}
Event reconstruction is important for any LArTPC data analysis. In this section, we describe in some detail the reconstruction chain that was used to derive the results reported in this paper. The first step in the reconstruction is to convert the raw signal from each wire to a standard ({\it e.g.} Gaussian) shape. This is achieved by passing the raw data through a calibrated deconvolution algorithm to remove the impact of the LArTPC electric field and electronics responses from the measured signal~\cite{Baller:2017ugz}.

The hit-finding algorithm scans the processed wire waveform looking for local minima. If a minimum is found, the algorithm follows the waveform after this point until it finds a local maximum. If the maximum is above a specified threshold, the program scans to the next local minimum and identifies this region as a hit. Hits are fit with a Gaussian function whose features identify the correct position (time coordinate), width and height and area (deposited charge) of the hit. A single Gaussian function is used to describe hits produced by isolated single particles. Multiple Gaussian functions are used to describe hits in a region where there are overlapping particles (e.g. around neutrino interaction vertex).

Clusters are reconstructed by the TrajCluster algorithm~\cite{Baller:2017ugz} as a collection of hits in the same wire plane that may be grouped together due to proximity to one another. TrajCluster incorporates elements of pattern recognition and Kalman filter fitting. The concept is to construct a short ``seed'' trajectory of nearby hits. Additional nearby hits are attached to the leading edge of the trajectory if they are similar to the hits already attached to it. The similarity requirements use the impact parameter between the projected trajectory position and the prospective hit, the hit width and the hit charge. This process continues until a stopping condition is met such as lack of hits, an abnormally high or low charge environment, encountering a two-dimensional (2D) vertex.

Three-dimensional (3D) tracks are constructed by the Projection Matching Algorithm (PMA)~\cite{Antonello:2012hu}. PMA builds and optimizes objects in 3D space by minimizing the cost function calculated simultaneously in all available 2D projections. In the configuration for the ArgoNeuT event reconstruction, PMA uses output from TrajCluster as input and attempts also to correct hit to cluster assignments using properties of 3D reconstructed objects. Vertices are also reconstructed and used to refine track reconstruction. A vertex can be the start point of an isolated track or the intersection point of several tracks originating from a common point.

\begin{figure*}[!ht]
\centering
\includegraphics[width=0.41\textwidth]{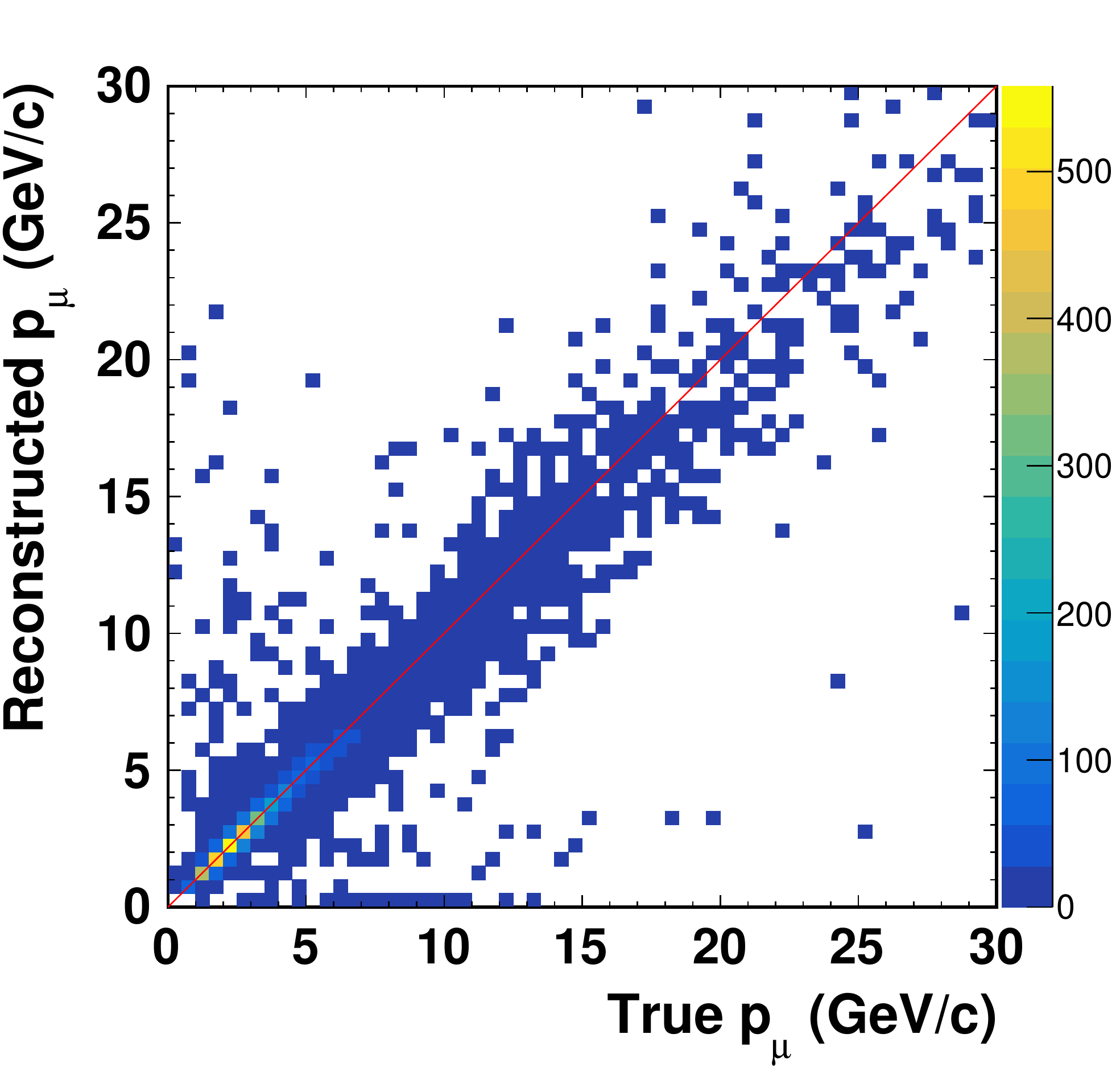}
\includegraphics[width=0.41\textwidth]{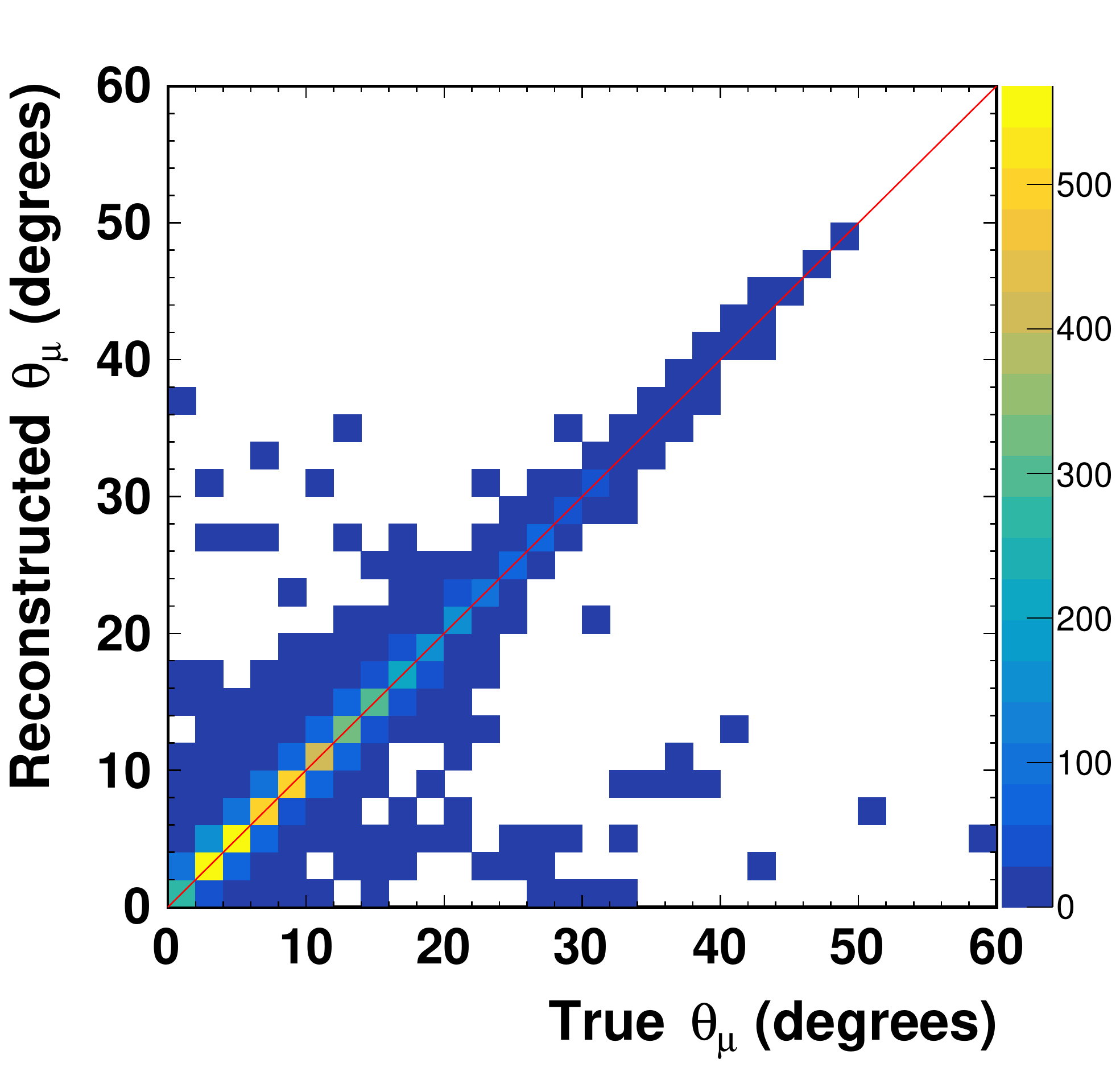}
\includegraphics[width=0.41\textwidth]{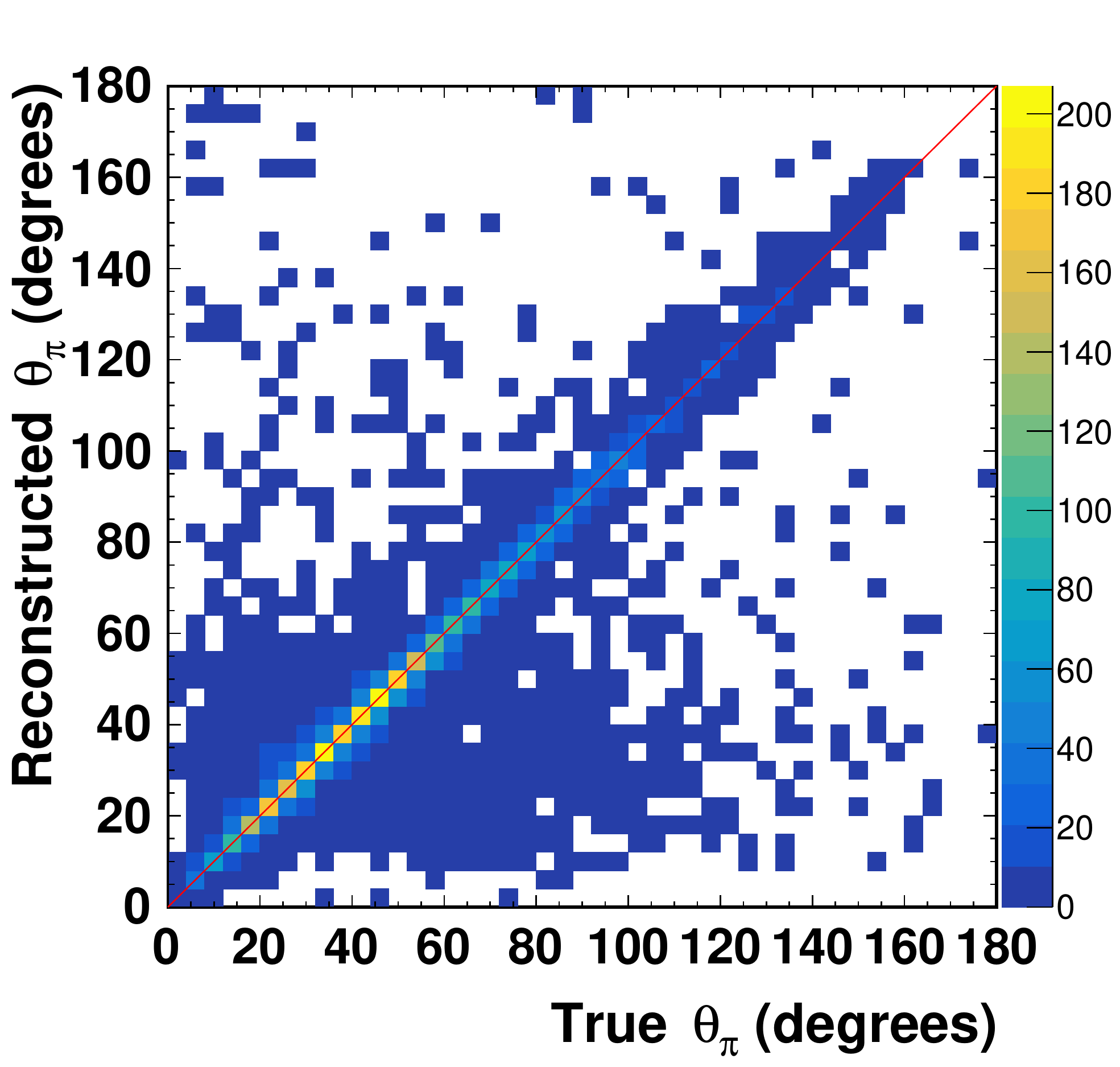}
\includegraphics[width=0.41\textwidth]{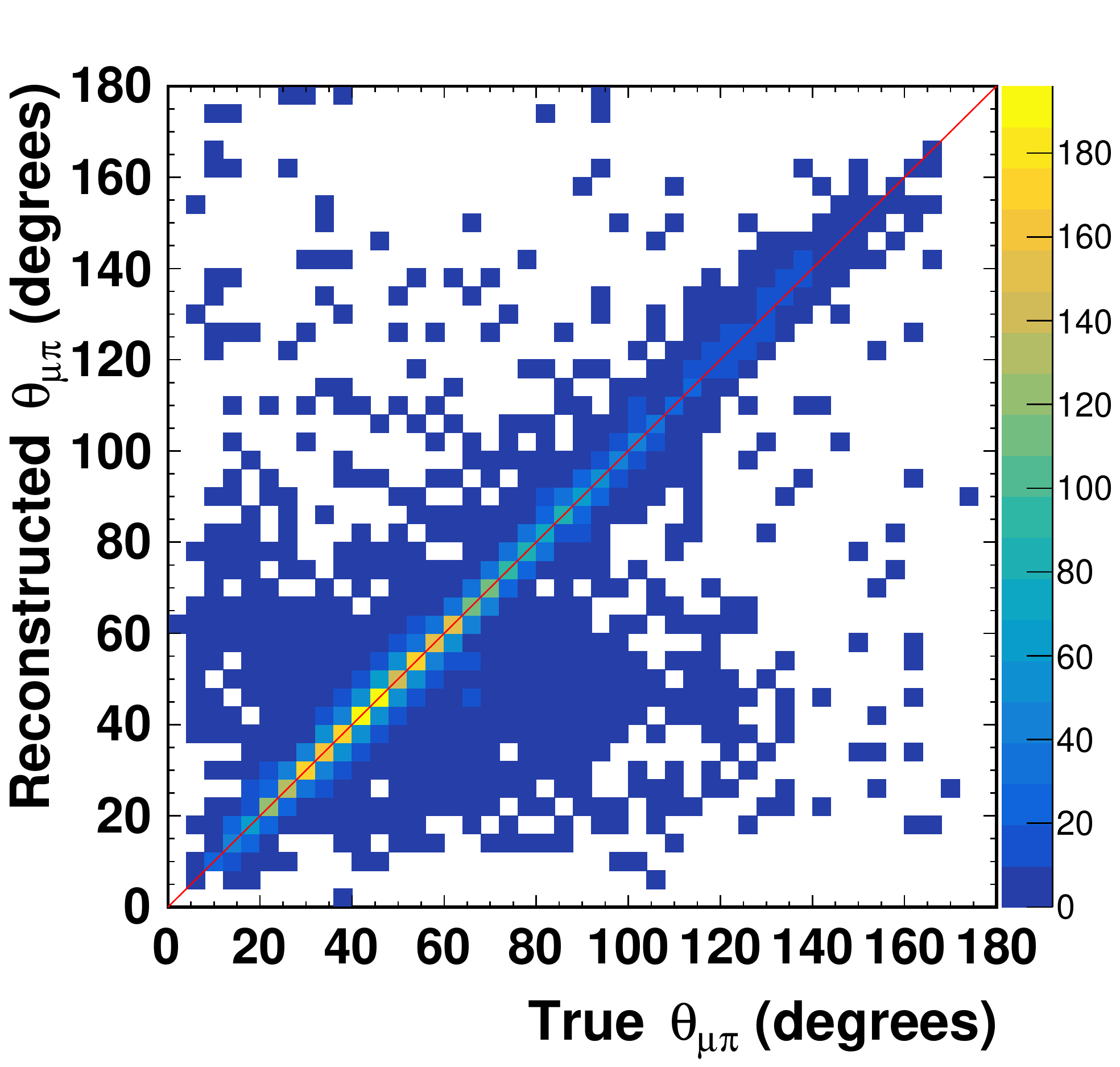}
\caption{Reconstructed quantity vs truth quantity in the selected $\nu_{\mu}$ and $\bar{\nu}_{\mu}$ samples. \commoncaption}\label{fig:resolution}
\end{figure*}

Figure~\ref{fig:resolution} shows the reconstructed quantity vs truth quantity for the four kinematic variables we measure in this analysis. The resolution on $p_{\mu}$ is 9\%. The resolutions on $\theta_{\mu}$, $\theta_{\pi}$ and $\theta_{\mu\pi}$ are $1^{\circ}$, $3^{\circ}$ and $3^{\circ}$. 

\begin{figure}[h]
\centering
\includegraphics[width=0.45\textwidth]{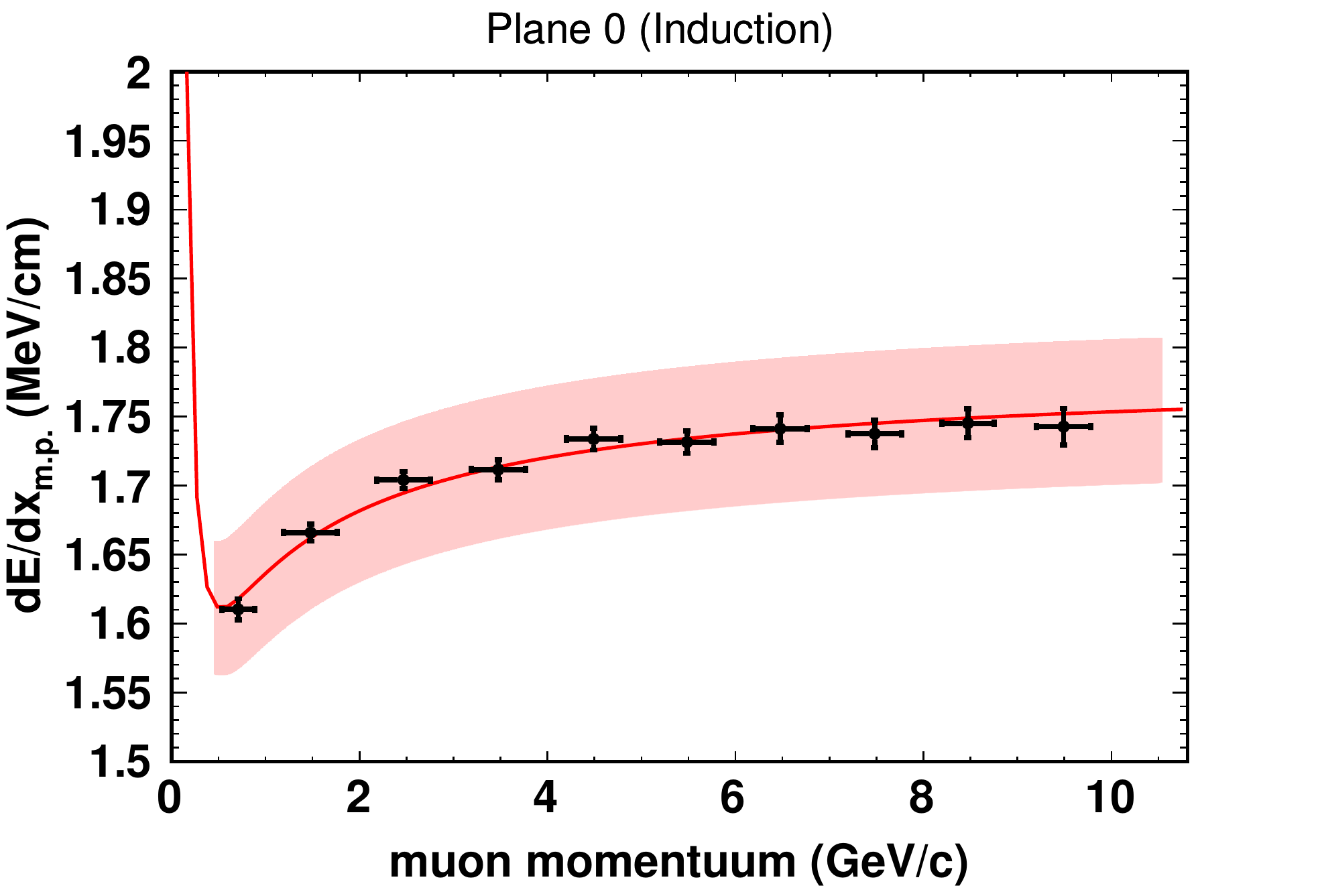}
\includegraphics[width=0.45\textwidth]{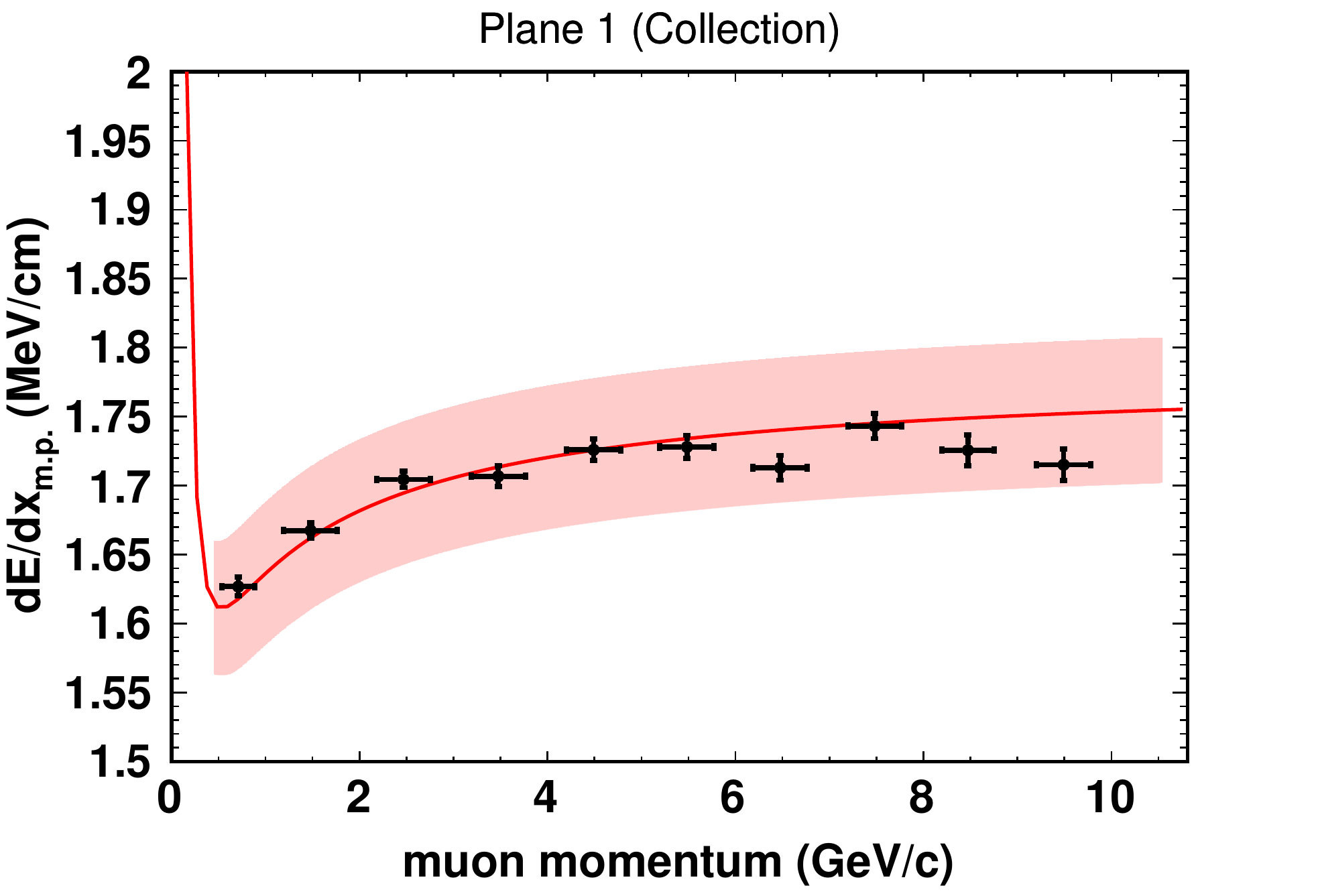}
\caption{Most probable (m.p.) value of $dE/dx$ in the last 5 cm of the matched muon tracks as a function of muon momentum measured by MINOS. The top figure is for the induction plane while the bottom figure is for the collection plane. The red curves are the {\sc geant} prediction for muons. The shaded area represent a 3\% uncertainty on the energy scale.}\label{fig:dedxmu}
\end{figure}

The measured pulse area (ADC) is converted to the number of electrons by an electronic calibration factor derived using through-going muons in the LArTPC. We select events where there is only one reconstructed track in the ArgoNeuT TPC. The track is required to match to a track in the MINOS detector. We calculate the most probable $dE/dx$ value using the last 5 cm of the track in argon. We plot the distribution of $dE/dx$ for different muon momenta. We tune the electric calibration factors so that the most probable value of $dE/dx$ as a function of muon momentum agree with the {\sc geant} prediction. Fig.~\ref{fig:dedxmu} shows the most probable value of $dE/dx$ as a function of muon momentum after the tuning of electronic calibration factors, which is in agreement with the {\sc geant} prediction shown as red curves. The shaped area represent a 3\% uncertainty on the energy scale, which is used in the systematic analysis.

 The charge deposition per unit length along the track direction ($dQ/dx$) is converted to $dE/dx$ after correcting for the impurities in the liquid argon and recombination effects~\cite{Acciarri:2013met}. If the incident particle stops in the LArTPC active volume, $dE/dx$ as a function of the residual range ($R$), the path length to the end point of the track, is used as a powerful method for particle identification. The software calculates four $\chi^{2}$ values for each track by comparing measured $dE/dx$ versus $R$ points to the proton, charged kaon, charged pion and muon hypotheses and identifies the track as the particle that gives the smallest $\chi^{2}$ value. However, because of the small size of ArgoNeuT, many particles are not contained in the TPC, which makes the particle identification a challenge. The average $dE/dx$ of exiting tracks still provides useful information for particle identification and is used in this analysis.

%\begin{figure}[h]
%\centering
%\includegraphics[width=0.45\textwidth]{dEdxrrp.pdf}
%\includegraphics[width=0.45\textwidth]{dedxrrpdata.pdf}
%\caption{$dE/dx$ as a function of residual range for identified contained proton tracks and charged pion tracks in the simulation. The red curves represent the expected average $dE/dx$ profiles for protons and charged pions.}\label{fig:dedxpro}
%\end{figure}

%\section{\label{sec:unfolding}Unfolding Procedure}
%This appendix discusses the resolution of the measured kinematic variables and the unfolding procedure. 

%As explained in Sec.~\ref{sec:eff}, the  bin-by-bin  unfolding  factor, $\epsilon_{\mathrm{UF}}$ in Eq.~\ref{eq:sigma}, combines corrections for acceptance, efficiencies of the event selection, and resolution effects. 

\end{document}